\documentclass[usenatbib]{mnras}

\usepackage{newtxtext,newtxmath}
\usepackage[T1]{fontenc}

\DeclareRobustCommand{\VAN}[3]{#2}
\let\VANthebibliography\thebibliography
\def\thebibliography{\DeclareRobustCommand{\VAN}[3]{##3}\VANthebibliography}

\usepackage{graphicx}	
\graphicspath{{Images/}}
\usepackage{amsmath}	
\usepackage[table,xcdraw]{xcolor}    
\usepackage{subfig}
\usepackage{comment}
\usepackage[normalem]{ulem}
\usepackage{physics}
\usepackage{xcolor}
\usepackage{lipsum}
\usepackage{xfrac}

\definecolor{royalazure}{rgb}{0.0, 0.22, 0.66}
\definecolor{auburn}{rgb}{0.43, 0.21, 0.1}
\definecolor{bostonuniversityred}{rgb}{0.8, 0.0, 0.0}
\definecolor{planet}{HTML}{69DF45}

\usepackage{hyperref}
\hypersetup{
    colorlinks=true,
    linkcolor=bostonuniversityred,
    linktoc=all,
    filecolor=blue,      
    urlcolor=royalazure ,
    citecolor=royalazure
}

\title[Leading \& Trailing Spirals in Protoplanetary Discs]{Leading \& Trailing Spiral Arms in a Nearly Broken Protoplanetary Disc}

\author[S. Rowther et al.]{Sahl Rowther,$^{1}$\thanks{E-mail: sahl.rowther@leicester.ac.uk} Rebecca Nealon,$^{2,3}$ Richard Alexander$^{1}$ and Farzana Meru$^{2,3}$
\\ 
$^{1}$School of Physics and Astronomy, University of Leicester, Leicester LE1 7RH, UK\\
$^2$Centre for Exoplanets and Habitability, University of Warwick, Coventry CV4 7AL, UK\\
$^{3}$Department of Physics, University of Warwick, Coventry CV4 7AL, UK
}

\date{Accepted XXX. Received YYY; in original form ZZZ}

\pubyear{2025}

\begin{document}
\label{firstpage}
\pagerange{\pageref{firstpage}--\pageref{lastpage}}
\maketitle 

\begin{abstract}
We perform three-dimensional smoothed particle hydrodynamics simulations to investigate the formation of spiral arms in misaligned circumbinary discs. In a nearly broken disc the misaligned inner and outer discs interact at two nodes, launching leading spiral arms that do not rotate with the disc. These spirals vanish when the disc is fully broken or aligned. Our results show that the formation of leading spirals is driven by the relative misalignment of the inner and outer disc, and does not depend on the disc physics. With live radiative transfer, the shadows cast by the misaligned inner disc are also able to launch trailing spiral arms that only appear at high misalignments when the discs are disconnected. When the disc is strongly misaligned, leading and trailing spiral arms can both appear and interact with each other. At lower misalignments, the impact of shadows is negligible and leading spiral arms are seen instead. The presence of both leading and trailing spiral arms implies that the rotation of the disc cannot be assumed based on the orientation of the spiral arms alone. Unlike spirals formed by gravitational instability, the spirals in this work can also form in low-mass, gravitationally stable discs.
\end{abstract}




\section{Introduction}

Recent high resolution observations of protoplanetary discs have revealed a variety of substructures \citep{2018Andrews,2018Long}. Common morphologies include ring \& gap structures \citep{2015ALMA,2018Huang,2018Sheehan,Segura-Cox2020}, large-scale spiral structures \citep{2015Benisty,2016Perez,2018bHuang,2021Paneque,2024Speedie}, and shadows \citep{2015Marino,2017Benisty,2020Keppler,2020Muro-Arena}.

The diversity of substructures also indicate a variety of mechanisms to explain their origins. Rings \& gaps are typically explained by planet-disc interactions \citep{2012Kley}. Large-scale spiral structures can be caused by gravitational instability or massive companions \citep{2017Meru,2018bForgan}. While these structures are easily explained by coplanar discs, the presence of shadows require a more complicated non-coplanar geometry.  Shadows are cast by an inner disc that is misaligned relative to the outer disc \citep{2018Facchini,2020Nealon,2021Ballabio}. Such discs are usually referred to as warped or broken discs depending on whether the disc is continuous or not. The presence of warps are also commonly seen as strong lateral asymmetries in 75\% of edge-on discs \citep{2024Villenave}. A high occurrence of shadowing in discs is also seen in \citet{2024Garufi}. Approximately 30\% of transition discs have also been found to be misaligned \citep{2022Bohn}.

There are multiple mechanisms that can warp a protoplanetary disc. These include a misaligned internal \citep{2013Facchini} or external companion \citep{2020aNealon}, or misaligned infall from the star-forming environment \citep{2018Bate,2019Sakai}. In addition to warping the disc, these mechanisms can form large-scale spirals due to interactions with a companion \citep{2023Smallwood} or trigger gravitational instability by infalling gas increasing the mass of the disc \citep{2025Speedie}. Warps are also able to suppress gravitational instabilities resulting in ring \& gap structure \citep{2022Rowther}. Additionally, the shadows in misaligned discs are also able to launch non-rotating spiral arms \citep{2016Montesinos,2019bCuello,2024Zhang,2024Su}. Warps, and the mechanisms that cause them, actively drive the evolution of discs and explain a range of observed structures. Thus, studying warped and broken discs is essential to understanding how protoplanetary discs are shaped. 

The influence of shadows on the disc dynamics has been modelled by \citet{2016Montesinos,2019bCuello,2024Zhang,2024Su} by prescribing azimuthal variations in the temperature profile. This leads to pressure gradients which excite trailing spiral arms. While these prescriptions for the temperature structure mimick the effect of a misaligned inner disc, these studies do not include the dynamics of the inner disc. By contrast, \cite{2020Nealon} did not find any spiral structures in simulations with live radiative transfer and an evolving inner disc structure. Using live radiative transfer naturally accounts for shadows. However, due to the physical and numerical differences between these studies, the role of both shadows and disc hydrodynamics in driving these spirals remains unclear.

We aim to investigate the impact of the warped disc geometry and shadows on the dynamical evolution of the disc. The 3D hydrodynamics simulations are described in \S\ref{sec:model}. We present our results in \S\ref{sec:results}. We compare with previous work and discuss observational implications in \S\ref{sec:Discussion}. We conclude our work in \S\ref{sec:conclusion}.

\section{Methods}
\label{sec:model}

\begin{figure*}
    \centering
    \includegraphics[width=\linewidth]{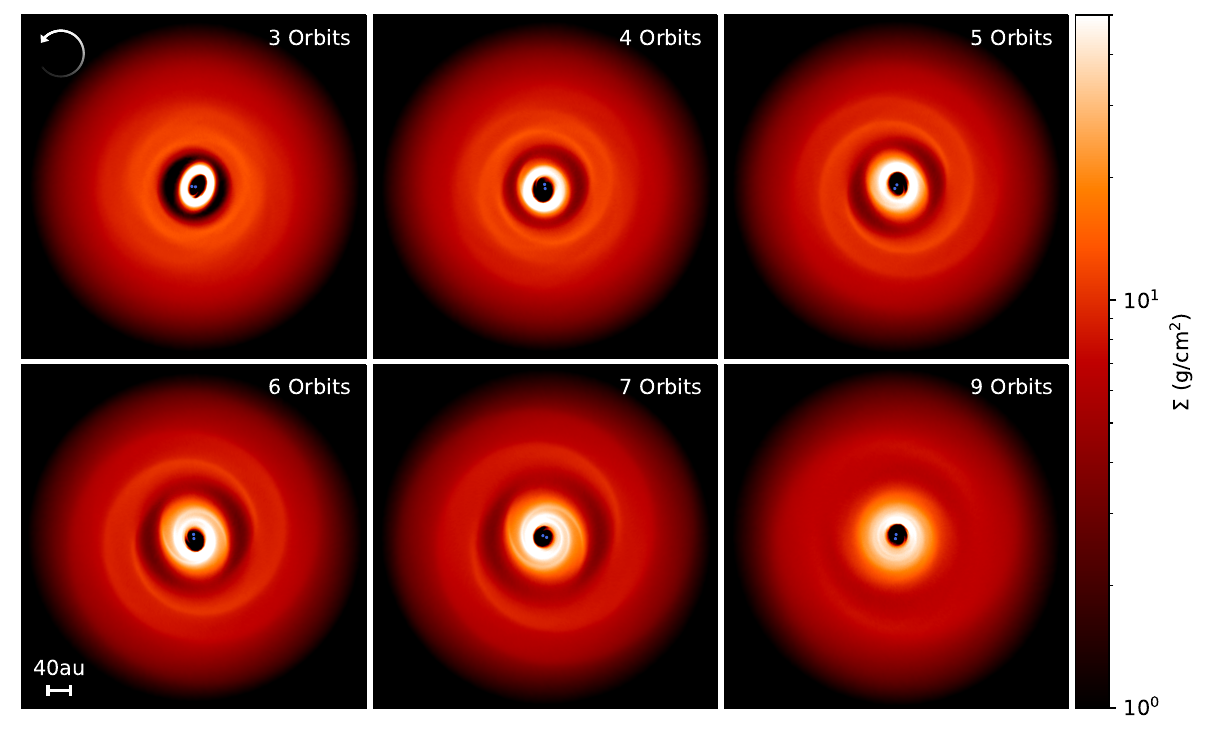}
    \caption{The evolution of the surface density of a $0.2M_\odot$ circumbinary disc. The orbital plane of the binary is initially misaligned by $45^\circ$ relative to the outer disc. Spiral arms are formed when the inner and outer disc are moderately misaligned relative to each other. Unlike typical spiral arms, these spirals are leading instead of trailing. Additionally, they do not rotate with the disc. When the disc is aligned or completely broken, the leading spirals cease to exist. The curved arrow shows the rotation of the disc (counter-clockwise).}
    \label{fig:isoEvol}
\end{figure*}

We use \textsc{Phantom}, a smoothed particle hydrodynamics (SPH) code developed by \cite{2018Price} for all simulations in this paper. A subset of simulations couple \textsc{Phantom} with \textsc{mcfost}, a Monte-Carlo radiative transfer code developed by \cite{2006Pinte,2009Pinte}. We perform 6 simulations in total, described as follows.

\subsection{Fiducial Disc Setup}

We modelled the disc using $10^6$ SPH particles between  ${R_{\mathrm{in}} = 10}$\,au and ${R_\mathrm{out} = 250}$\,au in a $0.2M_{\odot}$ disc around an equal-mass binary with a semi-major axis of ${a=8}$\,au. The stars are modelled as sink particles \citep{1995Bate} with a mass of $0.5M_\odot$ and an accretion radius of 3\,au. The initial binary inclination is set to $45^\circ$. We set up an initial surface density profile $\Sigma = \Sigma_{0}  \left( {R}/{R_{\mathrm{in}}} \right)^{-1} f_{s}$,
where ${\Sigma_{0} = 1.77 \times 10^2\,\mathrm{g\ cm}^{-2}}$,  and ${f_{s} = 1-\sqrt{R_\mathrm{in}/R}}$ smooths the surface density at the inner boundary of the disc. We use a vertically isothermal equation of state where the sound speed is given by $c_s = c_{s,\mathrm{in}} \left( {R}/{R_{\mathrm{in}}} \right)^{-0.25}$. We do not include the self-gravity of the disc. The initial disc aspect ratio $H/R_\mathrm{in} = 0.05$ sets $c_{s,\mathrm{in}}$. The disc is resolved with initial approximate mean smoothing length over disc scale height of ${<}h{>}/H$ = 0.2 at the disc midplane.

We use the \cite{2010Cullen} switch to detect any shocks that form and generate the correct shock dissipation depending on the proximity to the shock. Near the shock, the linear shock viscosity parameter dissipates kinetic energy as $\alpha_{\text{AV}}\to 1$. Further away from the shock, $\alpha_{\text{AV}} \to 0$. We set the quadratic artificial viscosity parameter $\beta_{\text{AV}}=2$ to prevent particle interpenetration \cite[][]{2015Nealon,2018Price}. The effective viscosity used here is a numerical tool for shock capturing, and is not directly analagous to a \citet{1973SS} $\alpha$.

\subsection{Resolution Test}

\cite{2015Nealon} demonstrated that disc breaking is impacted by numerical resolution. To ensure numerical artefacts do not dominate over the physical processes, we perform a resolution study by comparing the evolution of the fiducial disc to an identical simulation with $8\times10^6$ SPH particles. This doubles the resolution of the disc with initial ${<}h{>}/H$ = 0.1.

\subsection{Including Self-Gravity \& Live Radiative Transfer}

Although the analysis is simpler with a vertically isothermal and non-self-gravitating disc, it is not realistic. A $0.2M_\odot$ disc is massive enough for its self-gravity to be non-negligible. In self-gravitating discs, modelling the thermodynamics is critical for an accurate understanding of its evolution \citep{2024bRowther}. If the disc is massive enough, the self-gravity of the disc can drive gravitational instabilities which take the form of large-scale spiral structures. Gravitational instabilities could interfere with any other structures that form due to the disc being warped. Accurate temperatures are also important for warped discs due to the misaligned inner disc casting shadows. Hence, we perform two additional simulations to ensure the results are independent of the physics being modelled. The first compares the impact of the disc thermodynamics by using \textsc{mcfost} to simulate the fiducial disc with live radiative transfer. The second simulation includes both self-gravity and live radiative transfer. We use an adiabatic equation of state for these additional simulations. The disc setup is otherwise identical to the fiducial simulation.

The stellar luminosity is set using the sink particle mass and a 3\,Myr isochrone from \cite{2000Siess}. This corresponds to two stars with a luminosity of ${L_\star = 0.30L_\odot}$, temperature of ${T_\star = 3758K}$, and a stellar radius of ${R_\star = 1.31R_\odot}$. We include $P\mathrm{d}V$ work and shock heating as source terms. As these are gas-only simulations, for the radiative transfer we assume that the dust is perfectly coupled to the gas with a constant dust-to-gas mass ratio of 0.01. The dust grains are distributed between $0.03-1000\,\mu\mathrm{m}$ using 100 grain sizes and a power-law exponent of $\mathrm{d}n(s) \propto s^{-3.5}\mathrm{d}s$. We assume spherical, homogeneous, astrosilicate dust grains \citep{Weingartner2001}.

Each \textsc{mcfost} Voronoi cell corresponds to one SPH particle. The high disc mass requires a large number of photons to ensure even the most optically thick regions are well sampled. Since we model similar disc masses as in \cite{2024bRowther}, we illuminate the disc using $5 \times 10^9$ photons. The temperatures are updated every ${1/(2\sqrt{2}) \approx 0.354}$ of a binary orbit, i.e. every 7.96 years. The high frequency ensures accurate temperatures in all parts of the disc since the dynamical time is always longer than the time between temperature updates.

Stellar radiation is absorbed by the disc in the low density upper layers where the optical depth $\tau = 1$. This layer re-emits the photons, illuminating and warming the disc midplane \citep{1997Chiang}. Hence, for accurate modelling of the disc temperatures through radiative transfer, the upper layers are required to be sufficiently resolved. This condition is satisfied for the simulations in this work, since the $\tau = 1$ surface lies within 2 disc scale-heights of the midplane, where the disc is still well-resolved.

\subsection{A Lower Disc Mass} 

The $0.2M_\odot$ disc is massive enough that its angular momentum is comparable to that of the binary. In this regime, the evolution of the disc precession and alignment is impacted by the disc's own angular momentum \citep{2005King}. Using a lower mass disc will investigate the regime where the binary is the dominant influence on the disc evolution.

Hence, we also simulate two additional simulations of a $0.02M_\odot$ disc. All other disc and binary parameters are identical to the fiducial simulation. The first simulation uses a vertically isothermal equation of state. The second uses an adiabatic equation of state with live radiative transfer to determine the disc temperatures. However, we reduce the number of photons to $2 \times 10^8$ to be similar to previous studies of gravitationally stable discs \citep{2020Nealon,2022aBorchert,2022bBorchert}. Since the disc is not as optically thick as the $0.2M_\odot$ disc, accurate temperatures are still obtained with a lower number of photons. Both simulations also include the self-gravity of the disc.

\section{Results}
\label{sec:results}

\subsection{Disc Evolution}
\label{sec:highMass_disc}

Figure \ref{fig:isoEvol} shows the surface density evolution of the fiducial simulation, a $0.2M_\odot$ circumbinary disc misaligned relative to the binary. The misaligned binary causes the disc to become warped and eventually breaks as seen at 3 outer orbits in the top left panel of Fig \ref{fig:isoEvol}. While the disc is broken, no spiral structures are visible. 
While the disc is realigning during the intermediate phase of the disc's evolution, leading spirals are formed as seen in the middle panels at ${4-7}$ outer orbits. These spirals differ from more typical mechanisms for forming spirals such as gravitational instability or planets in two important ways. Firstly, the spirals are pointing in the opposite direction, they are leading instead of trailing. Secondly, they do not rotate with the disc. The direction of rotation of the disc is represented by the curved arrow. The leading spirals disappear when the entire disc is in alignment with the binary and there is no longer an inner and outer disc as seen at 9 outer orbits in the bottom right panel of Fig \ref{fig:isoEvol}.

\begin{figure*}
    \centering
    \includegraphics[width=\linewidth]{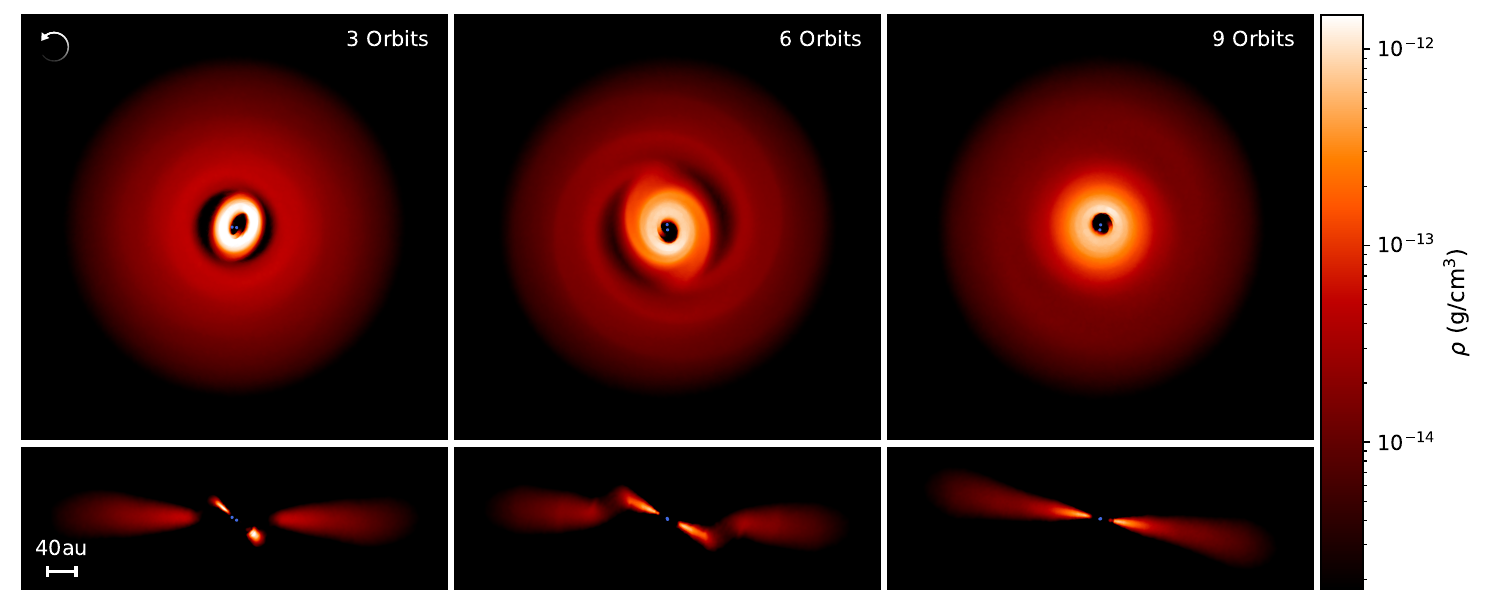}
    \caption{The top panels are particle plots of the density evolution. The bottom panels are cross-sectional slices of the density in the edge-on view of the disc. These three snapshots show the three scenarios that can occur depending on the relative disc misalignment between the inner and outer disc. When there is a strong relative disc misalignment, the disc is fully broken and there is no connection between the inner and outer disc (left panels). A moderate relative disc misalignment forms two leading spiral arms that originate where the inner and outer disc connect (middle panels). When the inner and outer disc have realigned, the spirals disappear since the inner and outer disc are no longer connected at two narrow nodes (right panels).}
    \label{fig:isoAltView}
\end{figure*}

The origin of the leading spirals is better seen in Figure \ref{fig:isoAltView} which shows the three scenarios that occur during the disc's evolution. The top panels are particle plots of the density evolution after the particles have been sorted in ascending order based on their density. Ordering the particles ensures the most visible particles are the ones with the highest density. These are the particles of interest since the spirals are regions of higher density compared to the background. At 3 and 9 outer orbits (the top left and top right panels, respectively) are similar to their counterparts in Figure \ref{fig:isoEvol} with no leading spirals visible when the disc is completely broken or aligned. However, at 6 outer orbits (top middle panel), the connection between the inner and outer disc is now clearly visible. The leading spirals originate at the locations where the inner and outer disc are connected. The locations of the two connecting nodes do not rotate with disc. Hence, the leading spirals do not rotate with the disc.

\begin{figure}
    \centering
    \includegraphics[width=\linewidth]{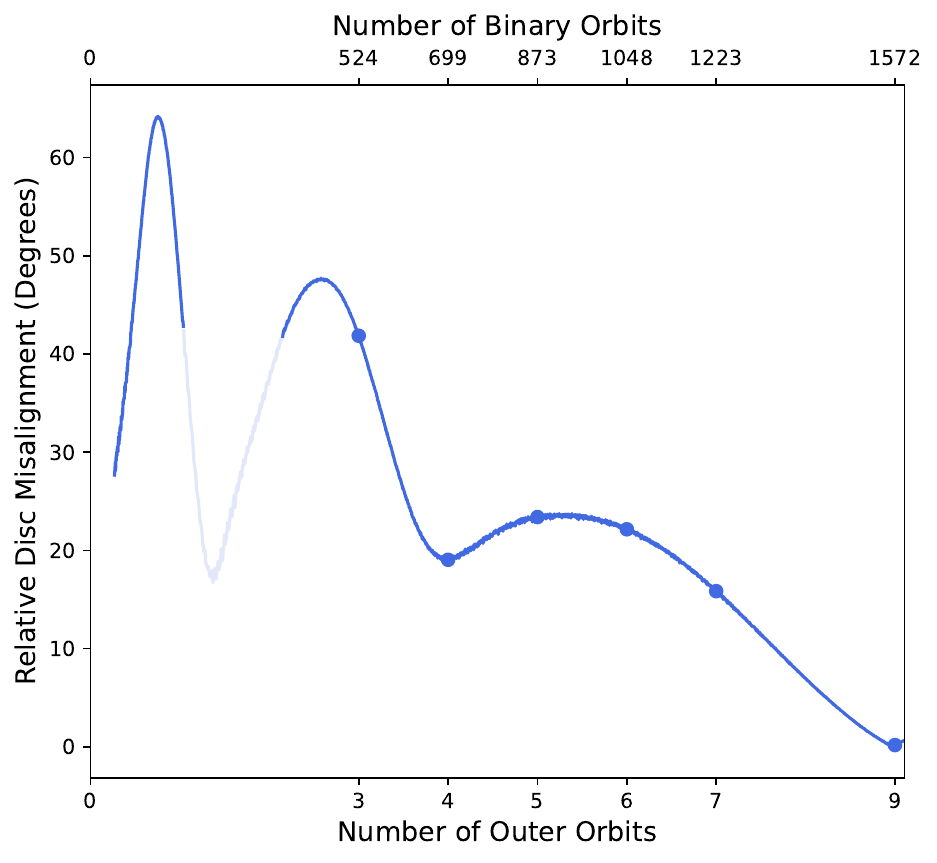}
    \caption{Time evolution of the relative disc misalignment $\theta$, between the inner and outer disc for the fiducial simulation. The markers represent the snapshots plotted in Figure \ref{fig:isoEvol}. The semi-transparent line shows when the inner disc is impacted by transient features between $1-2$ orbits. Leading spirals only form when ${10^\circ \lesssim \theta \lesssim 40^\circ}$ between $4-7$ orbits.}
    \label{fig:theta}
\end{figure}

\begin{figure*}
    \centering
    \includegraphics[width=\linewidth]{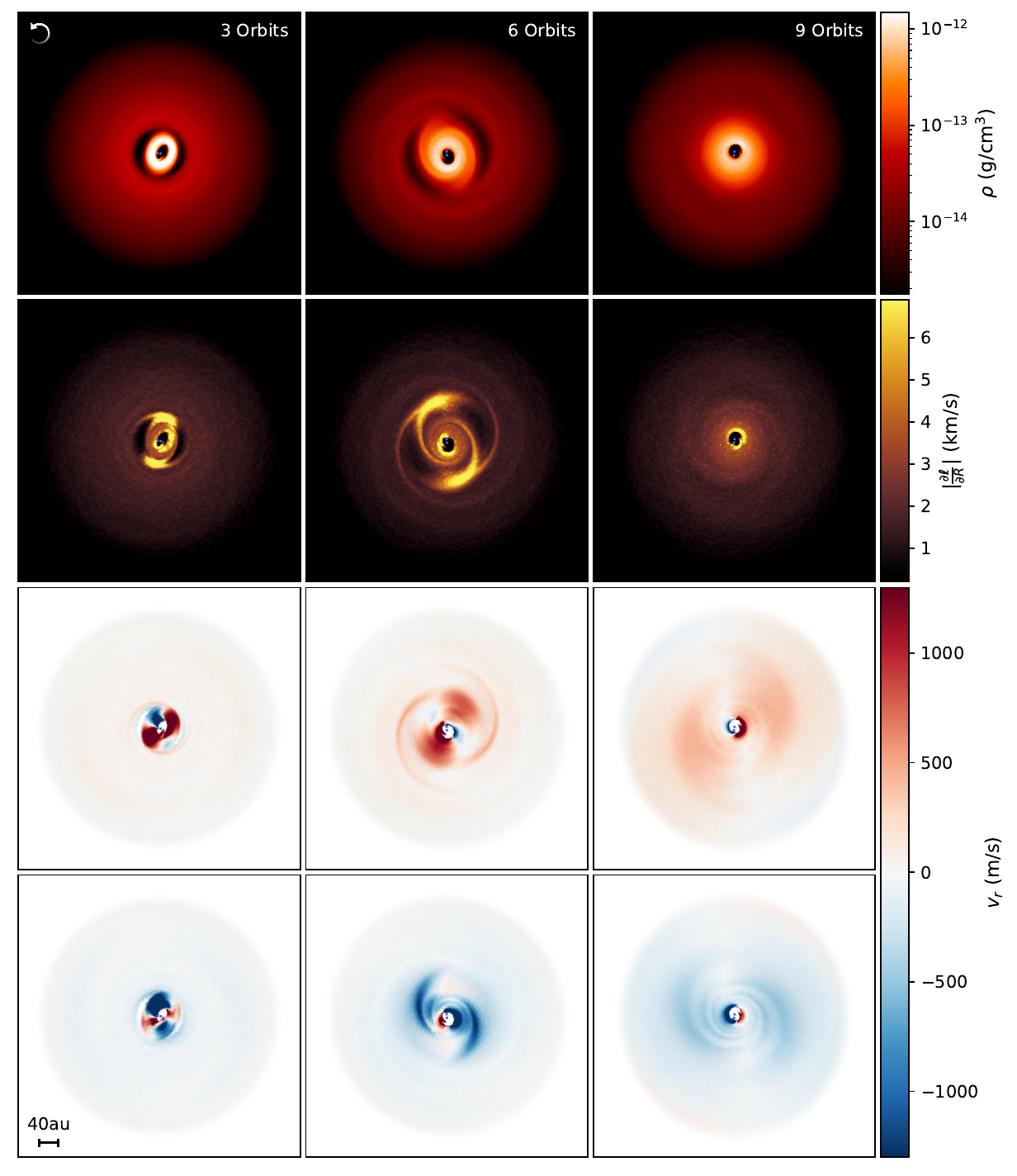}
    \caption{The first row shows particle plots of the density evolution. The second rows shows particle plots of the angular momentum transfer represented by the magnitude of the radial gradient of the specific angular momentum vector $\left|\partial\vb*{\ell}/\partial R\right|$.  The bottom rows are particle plots of the radial velocity, $v_r = v_x (x/R) + v_y (y/R)$. The panels in the third and fourth row show particles with outward and inward radial motion, respectively. These three snapshots show the three scenarios that can occur depending on the relative disc misalignment between the inner and outer disc. The leading spirals form due to the increased radial velocities driven by angular momentum transfer from a non-corotating source. The spirals being visible in all columns also demonstrate that they are not a visualisation or viewing effect of the density.}
    \label{fig:vr}
\end{figure*}

\begin{figure*}
    \centering
    \includegraphics[width=\linewidth]{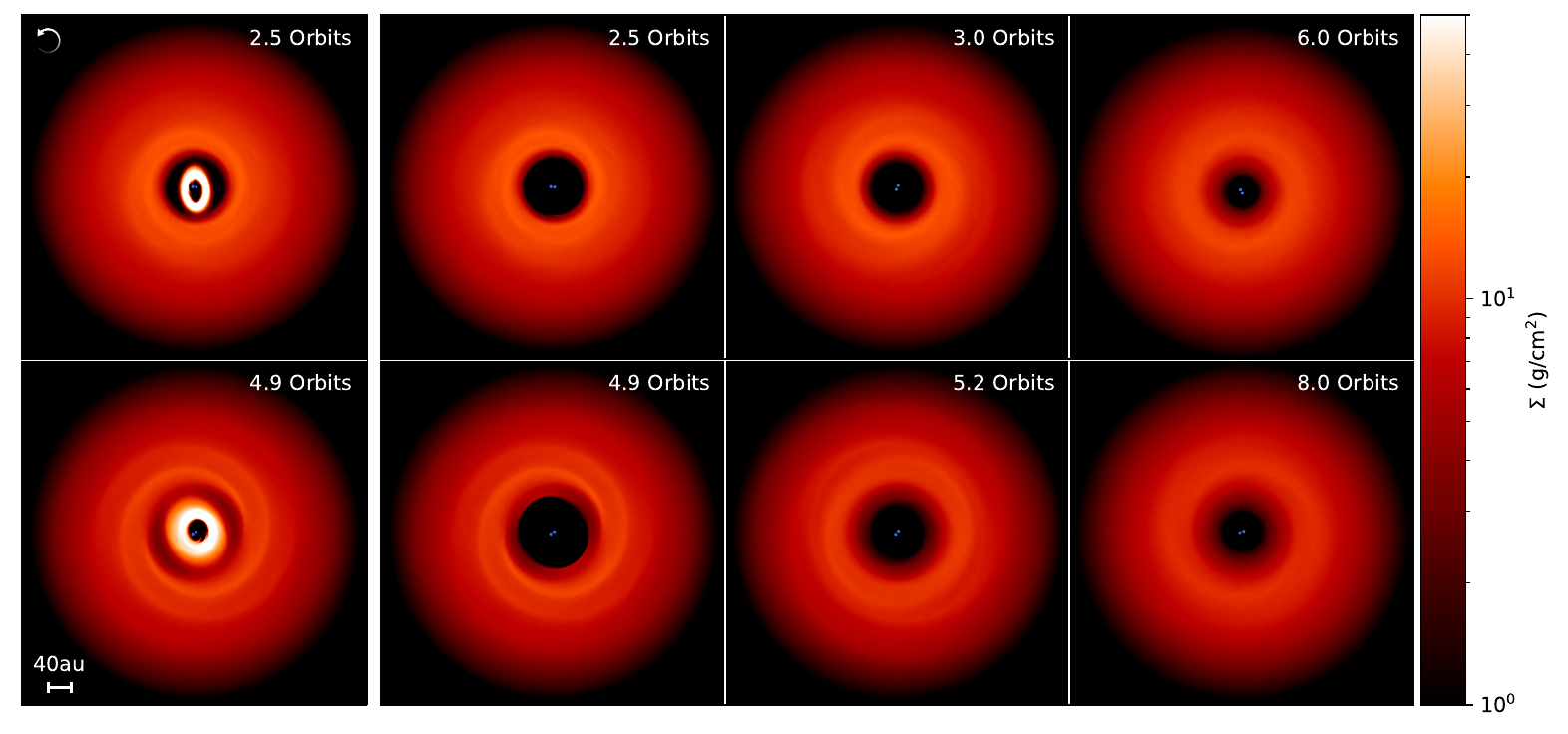}
    \caption{The left-most detached panels shows the surface density of the fiducial disc before and after leading spirals have formed at 2.5 and 4.9 orbits, respectively. The other three panels show the surface density evolution without the inner disc. When the inner disc is removed, the leading spiral arms cannot be formed or sustained. The lack of spirals in the right panels is evident when compared to Figure \ref{fig:isoEvol} where leading spiral arms are clearly visible.}
    \label{fig:innerDisc}
\end{figure*}

\begin{figure}
    \centering
    \includegraphics[width=\linewidth]{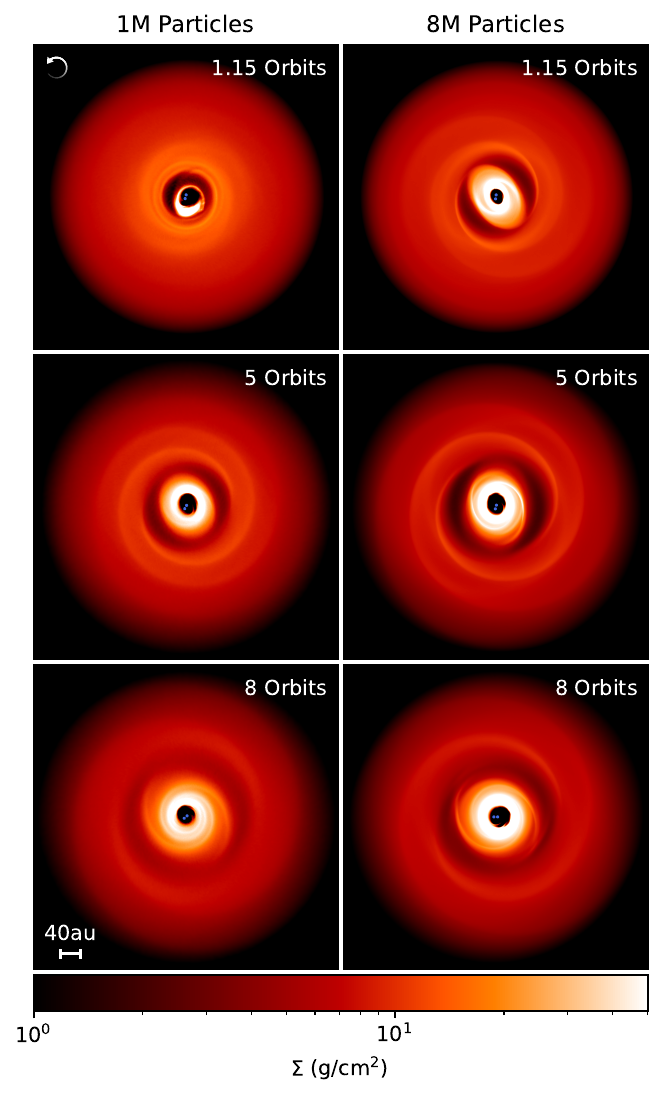}
    \caption{A resolution study comparing the impact of numerical resolution on the formation and evolution of leading spiral arms. The only major difference between the two simulations occurs at early times. At lower resolution, the inner disc is impacted by transient features that prevent leading spirals from forming (top left panel). The transient features are not present at higher resolution, and hence leading spirals are also seen at early times (top right panel). After the transient features have disappeared, the evolution of the leading spirals is consistent with both 1 and 8 million particles as seen in the middle and bottom panels.}
    \label{fig:resolution}
\end{figure}

\begin{figure}
    \centering
    \includegraphics[width=\linewidth]{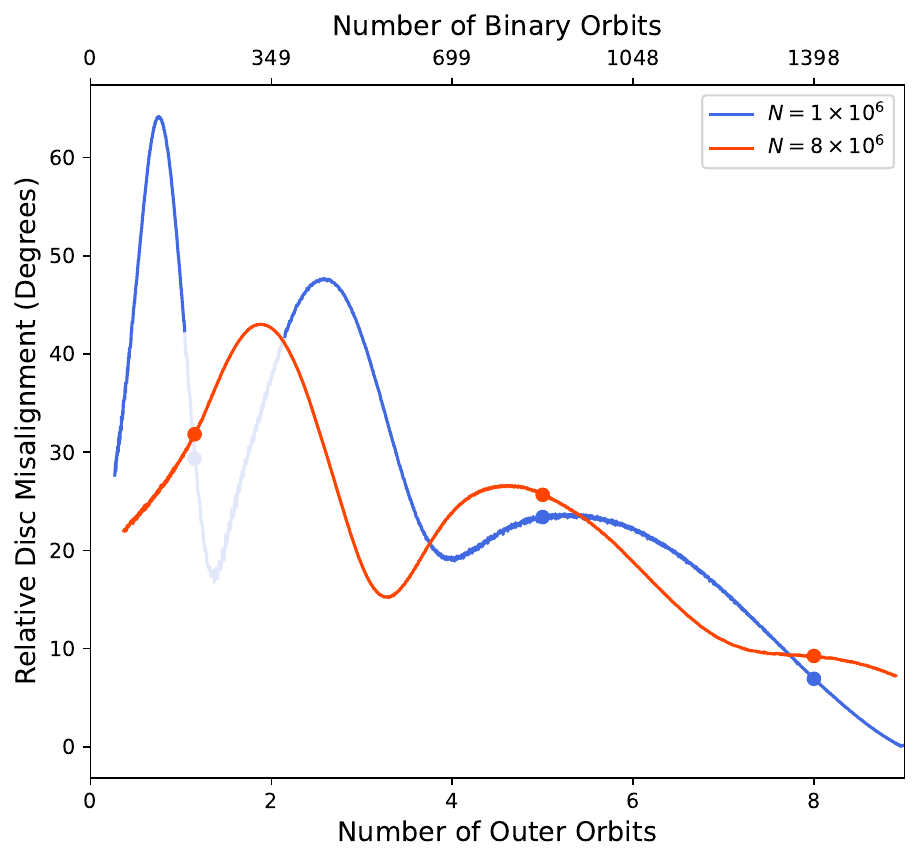}
    \caption{A comparison of the evolution of the relative disc misalignment at different resolutions. The markers represent the snapshots plotted in Figure \ref{fig:resolution}. The semi-transparent line shows when the inner disc is impacted by transient features between $1-2$ orbits. After the transient features have disappeared, the relative disc misalignment in both simulations evolves similarly. Thus, the formation and evolution of the leading spiral arms is not impacted by numerical resolution. }
    \label{fig:resolution_misalignment}
\end{figure}

\begin{figure*}
    \centering
    \includegraphics[width=\linewidth]{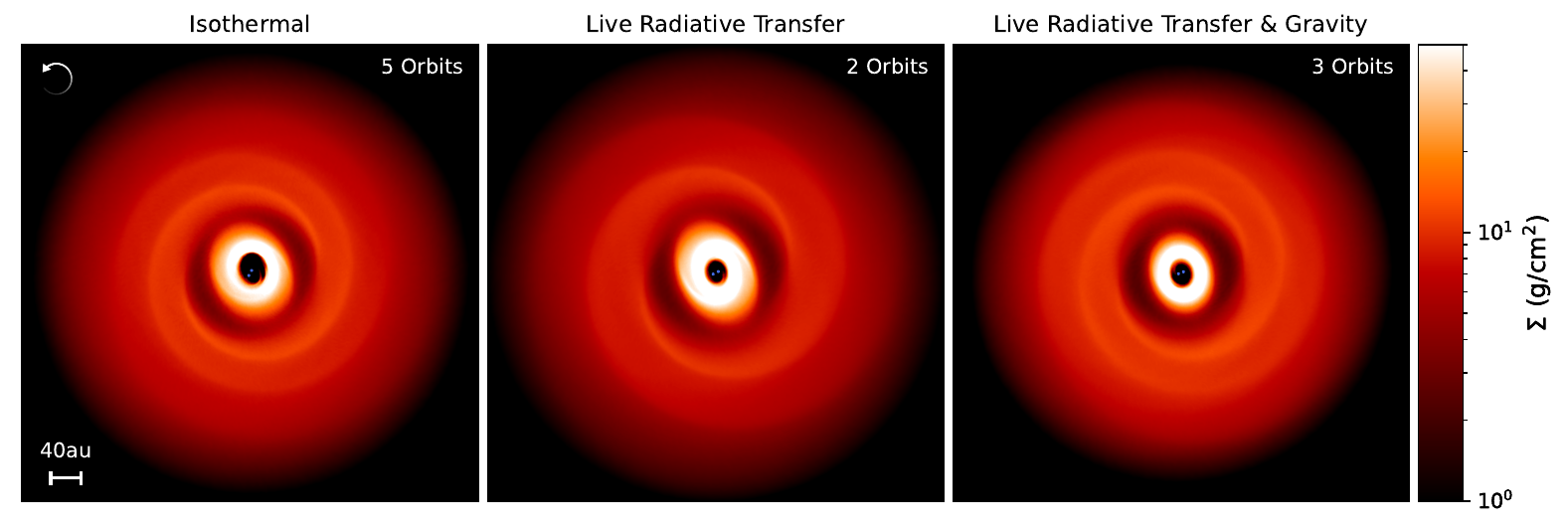}
    \caption{A comparison of the evolution of the $0.2M_\odot$ disc with different physics. The left panel shows the fiducial simulation modelled with a vertically isothermal equation of state and without self-gravity. The middle panel shows an identical disc modelled with live radiative transfer and an adiabatic equation of state. The right panel is identical to the middle panel, but also includes self-gravity. Despite the different physics, the formation of leading spiral arms is consistent across all three simulations.}
    \label{fig:physics}
\end{figure*}

The bottom panels show the cross-section slices of the density in the edge-on view of the disc. These panels show how the relative misalignment between the inner and outer disc determines the morphology of the disc. At 3 orbits, the inner and outer disc are strongly misaligned. This results in the disc to be completely broken with the inner and outer disc being disconnected from each other. At 9 orbits, there is no longer any misalignment between the inner and outer discs. Leading spirals are only visible when there is a moderate misalignment between the inner and outer disc. This is the case at ${\sim}\,6$ orbits. As seen in the middle panel, the moderate misalignment allows the inner disc to still be connected with the outer disc. The leading spirals originate at these two connecting nodes.

The relative disc misalignment $\theta$, is calculated using the dot product of the unit angular momentum vector $\vb*{\hat{\ell}}$,  of the inner and outer disc,
\begin{equation}
    \theta = \cos^{-1} \left( \vb*{\hat{\ell}}_\mathrm{inner} \cdot \vb*{\hat{\ell}}_\mathrm{outer} \right).
\end{equation}
The location of the gap determines the inner and outer parts of the disc. Figure \ref{fig:theta} shows the evolution of the relative disc misalignment over time. The markers represent the snapshots shown in Figure \ref{fig:isoEvol}. From Fig \ref{fig:theta}, it is clear the leading spirals are only able to form when the inner and outer disc is moderately misaligned relative to each other, when $10^\circ \lesssim \theta \lesssim 40^\circ$. Between $4-7$ orbits, $\theta$ is consistently in this range. Comparing with Fig \ref{fig:isoEvol}, this is the only phase where the leading spirals are visible. At higher values of $\theta$, the disc is broken as seen at 3 orbits. At lower values, the inner and outer disc begins to fully reconnect. At these extremes, the inner and outer disc are no longer connected at two narrow nodes, and thus the spirals disappear. There is also a phase between $1-2$ orbits where the relative disc misalignment is in the right range for leading spirals to form. However, in this phase represented by the semi-transparent line, the inner disc is impacted by transient features and prevents the formation of leading spirals.

\subsection{Why are there Leading Spirals?}

The first row in Figure \ref{fig:vr} shows particle plots of the density of the three scenarios described above and are identical to the top panels in Figure \ref{fig:isoAltView}. The second row shows particle plots of angular momentum transfer, as a proxy for the torque, represented by the magnitude of the radial gradient of the specific angular momentum vector $\left|\partial\vb*{\ell}/\partial R\right|$. Since aziumthally averaging would erase any asymmetric features from the spirals, we instead calculated $\left|\partial\vb*{\ell}/\partial R\right|$ for each particle individually as
\begin{equation}
    \label{eq:dLdR}
    \left(\frac{\partial\vb*{\ell}}{\partial R}\right)_i = (\nabla \vb*{\ell})_i \cdot \hat{\vb*{r}}_i
\end{equation}
where we have projected the gradient of the angular momentum vector, $(\nabla \vb*{\ell})_i$, onto the radial direction using the spherical radial unit vector $\hat{\vb*{r}}_i$. Equation \ref{eq:dLdR} follows from the general expression of the gradient in spherical coordinates,
\begin{equation}
    \nabla \vb*{\ell} = \frac{\partial \vb*{\ell}}{\partial R}\ \hat{\vb*{r}} + \frac{1}{r} \frac{\partial \vb*{\ell}}{\partial \theta}\ \hat{\vb*{\theta}} + \frac{1}{r \sin \theta} \frac{\partial \vb*{\ell}}{\partial \phi}\ \hat{\vb*{\phi}}.
\end{equation}
For our SPH simulations, the gradient of the angular momentum vector is calculated in Cartesian coordinates using
\begin{align}
    (\nabla \vb*{\ell})_i &= \frac{1}{\Omega_i} \sum_j m_j \left(\frac{\vb*{\ell}_j - \vb*{\ell}_i}{\rho_j}\right) \otimes \nabla_i W_{ij}(h_i) \\
     &= \begin{bmatrix}
            \partial \vb*{\ell}_x / \partial x & \partial \vb*{\ell}_x / \partial y & \partial \vb*{\ell}_x / \partial z \\
            \partial \vb*{\ell}_y / \partial x & \partial \vb*{\ell}_y / \partial y & \partial \vb*{\ell}_y / \partial z \\
            \partial \vb*{\ell}_z / \partial x & \partial \vb*{\ell}_z / \partial y & \partial \vb*{\ell}_z / \partial z
        \end{bmatrix}_i
\end{align}
where $i$ and $j$ are particle labels, $m$, $h$, $\rho$, and $\vb*{\ell}$ represent the mass, smoothing length, density, and specific angular momentum vector, respectively. The term $\nabla_i W_{ij}(h_i) = \mathrm{d}W / \mathrm{d}r_{ij} \cdot \hat{\vb*{r}}_{ij}$ is the gradient of the smoothing kernel, and $\Omega_i$ accounts for a variable smoothing length and is given by
\begin{equation}
    \Omega_i = 1 + \frac{3h_i}{\rho_i} \sum_j m_j \frac{\partial W_{ij}(h_i)}{\partial h_i}.
\end{equation} 
To visualise where angular momentum transfer is at its strongest, we sort the particles based on $\left|\partial\vb*{\ell}/\partial R\right|$. Additionally, the particles have a transparency value based on their density. The bottom panels show the particle plots of the radial velocity calculated using the position and velocity of the particles as $v_r = v_x (x/R) + v_y (y/R)$. As before, the particles have a transparency value based on their density. In the third row, the particles have been sorted in ascending order based on their radial velocity to show particles with outward radial motion. In the fourth row, the particles have been sorted in descending order to show particles with inward radial motion.

Comparing the three scenarios in Figure \ref{fig:vr} reveals the key physics required for the formation of leading spiral arms. When the disc is misaligned, the torques are non-zero due to neighbouring annuli of gas having misaligned angular momentum vectors \citep{2007Lodato}. The torques result in angular momentum transfer across radii, exciting radial velocities that transport mass radially to redistribute angular momentum. At 6 orbits, the disc has a moderate relative disc misalignment, allowing the inner and outer discs to remain connected. Angular momentum can only be redistributed through these two narrow nodes. It is precisely in these regions where the leading and trailing spiral arms originate in the outer and inner disc, respectively. Outward and inward radial velocities are excited at these nodes resulting in the spiral structures seen in the density plots. At 3 orbits, the disc has a larger relative disc misalignment. The connection between the inner and outer disc is much weaker. Hence, the spirals are also weaker since angular momentum transfer between the inner and outer disc is less efficient. At 9 orbits, the inner and outer disc have realigned. Hence, the spirals disappear since the torques that drive angular momentum transfer are reduced. Additionally, any angular momentum transfer can now occur at all azimuths since the inner and outer disc are no longer connected at two narrow nodes.

Other mechanisms such as planet-disc interactions can also increase the radial velocities in a disc, but only trailing spirals are formed. A crucial difference from the mechanism in this work is that the source of the spirals (the planet) corotates with the disc. In contrast, the two connecting nodes do not corotate with the disc. The increased radial velocities due to angular momentum transfer from a non-corotating perturbation naturally results in leading spiral arms.

\subsection{The Role of the Inner Disc}

The importance of the inner disc is shown in Figure \ref{fig:innerDisc} which shows the evolution of the fiducial simulation after the inner disc has been removed. The detached panels show the surface density of the fiducial disc before and after leading spirals have formed at 2.5 and 4.9 orbits, respectively. The other panels on the right show the evolution of the disc without the inner disc. The top panels show that the removal of the inner disc prevents the leading spirals from ever forming. The bottom panels show that leading spirals disappear very quickly once the inner disc has been removed. Thus, it is clear that the connection to the inner disc is vital to forming and sustaining the spirals in Figure \ref{fig:isoEvol}.

\subsection{Resolution Test}

Since the formation of leading spirals depends on whether or not the disc is fully broken, a resolution test is necessary. Figure \ref{fig:resolution} compares the evolution of the fiducial disc setup at different resolutions. The left and right panels show the surface density evolution with $1 \times 10^6$ and $8 \times 10^6$ particles, respectively. Figure \ref{fig:resolution_misalignment} compares the evolution of the relative disc misalignment $\theta$, between the two simulations. Three snapshots are chosen to highlight the similarities (at 5 and 8 orbits) and the differences (at 1.15 orbits) at different resolutions.

The only major difference between the two simulations occurs in the first couple of orbits. At lower resolution, the inner disc shows transient features (top left panel of Fig \ref{fig:resolution}) which disappear as the disc relaxes into a more stable state. However, at higher resolution the inner disc is more stable. Thus, leading spirals are seen at earlier times at high resolution. 

After the transient features have disappeared, the evolution of the relative disc misalignment is similar for both simulations, with moderate misalignments, ${10^\circ \lesssim \theta \lesssim 40^\circ}$, between $4-7$ orbits (see Fig \ref{fig:resolution_misalignment}). During this time, both simulations have clearly visible leading spiral arms (middle panels of Fig \ref{fig:resolution}). At 8 orbits, the spirals are a little stronger in the higher resolution simulation. This is due to small differences in the relative disc misalignment at different resolutions. The stronger spirals at 8 orbits at higher resolution are due to larger relative disc misalignments. These differences are minor. Both simulations also evolve similarly in the final phase of the disc's evolution when the inner and outer disc is realigning. The leading spiral arms gradually weaken in this phase when $\theta \lesssim 10^\circ$ (bottom panels of Fig \ref{fig:resolution}). 

The consistent results from this resolution study confirm that the numerical resolution does not impact the formation of leading spiral arms at the connecting nodes of the inner and outer disc.

\subsection{Leading Spirals are Independent of Disc Physics}
\label{sec:physics}

Figure \ref{fig:physics} compares the evolution of the fiducial disc setup with different physics.  The left panel shows the vertically isothermal simulation without self-gravity described in \S\ref{sec:highMass_disc}. The simulation in the middle panel uses live radiative transfer to calculate the disc temperatures. The final simulation in the right panel also includes self-gravity, in addition to using live radiative transfer. The different physics leads to differences in the disc evolution. Hence, we plot the 3 simulations when their relative disc misalignment is similar (between $25-30^\circ$).

Using live radiative transfer results in a calculated temperature profile where there are temperature differences due to shadows cast by the misaligned inner disc.  While there are some differences in the disc evolution, the shadows have a negligible impact on the formation of the leading spiral arms. Their appearance remains consistent across all simulations. The leading spirals always form when the relative disc misalignment is between $10 - 40^\circ$. The inclusion of self-gravity demonstrates that the leading spiral arms overcome any gravitational instabilities that would have formed trailing spirals in an aligned disc. The impact of the misalignment on the spiral structure is greater than the impact of the disc self-gravity.

\subsection{The Evolution of a Lower Mass Disc}

\subsubsection{A Vertically Isothermal Disc}

\begin{figure*}
    \centering
    \includegraphics[width=\linewidth]{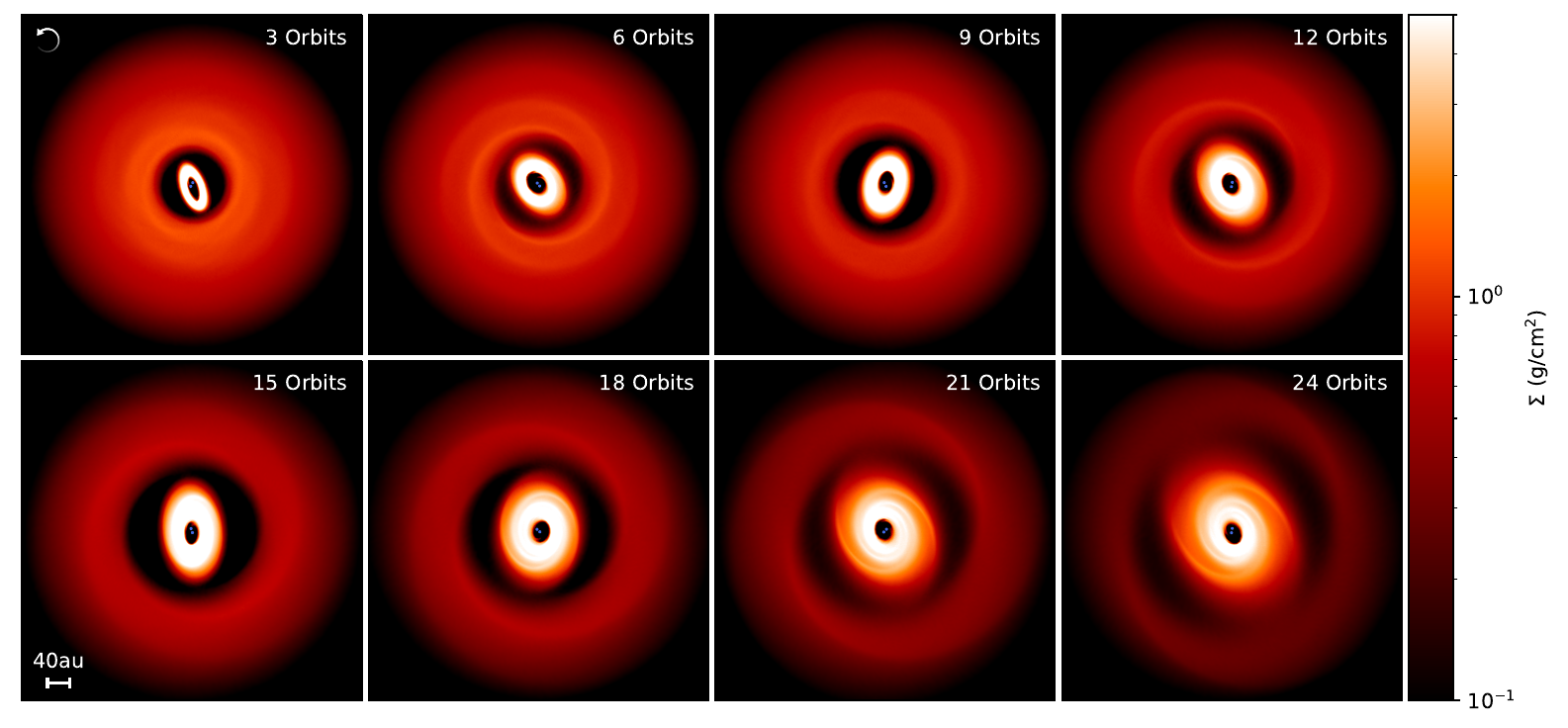}
    \caption{The evolution of the surface density of a $0.02M_\odot$ circumbinary disc. The orbital plane of the binary is initially misaligned by $45^\circ$ relative to the outer disc. Leading spiral arms are also formed in the lower mass disc. However, unlike the $0.2M_\odot$ disc, the spirals do not have a continuous presence. Instead, the disc structure alternates between clearly visible leading spirals, and where they have weakened over time.}
    \label{fig:lowMass}
\end{figure*}

Figure \ref{fig:lowMass} shows the surface density evolution of a $0.02M_\odot$ circumbinary disc misaligned relative to the binary every 3 outer orbits between 3 and 24 orbits. Similar to its higher mass counterpart in \S\ref{sec:highMass_disc}, leading spiral arms are formed due to relative misalignment between the inner and outer disc. However, the evolution of the spirals is different. In the $0.02M_\odot$ disc, the spirals do not have a continuous presence after they form as in the $0.2M_\odot$ disc. Instead, they alternate between being clearly visible, and where they have weakened over time.  

The difference in the evolution of the leading spiral arms arises because the alignment \& precession of the disc is no longer the same in the lower mass disc. 
In the $0.2M_\odot$ disc, the angular momentum of the disc is non-negligible and contributes to the total angular momentum vector of the system. Therefore, the disc and binary align towards each other resulting in the rapid alignment seen in Figure \ref{fig:theta}. 

\begin{figure}
    \centering
    \includegraphics[width=\linewidth]{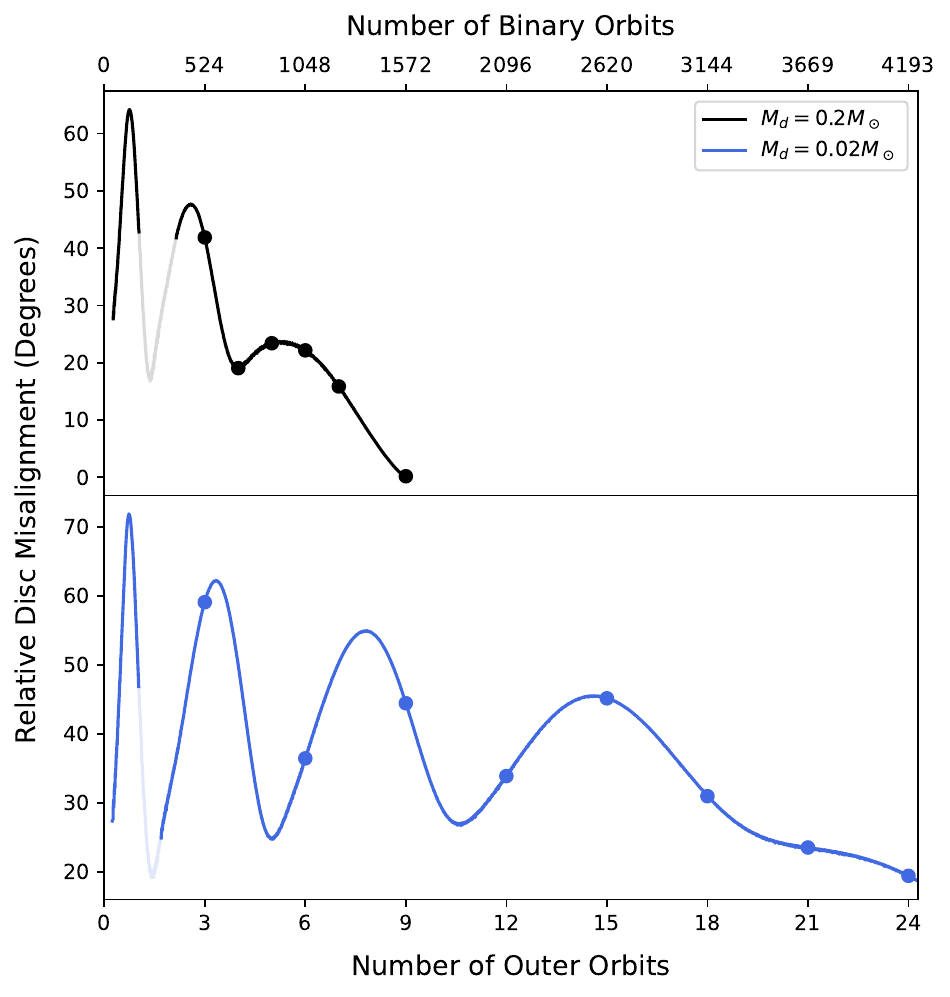}
    \caption{Time evolution of the relative disc misalignment $\theta$, between the inner and outer disc for the $0.2M_\odot$ (top panel) and the $0.02M_\odot$ disc (bottom panel). The markers represent the snapshots plotted in Figures \ref{fig:lowMass} and \ref{fig:theta}. The semi-transparent line shows when the inner disc is impacted by transient features between $1-2$ orbits. Leading spirals form when ${10^\circ \lesssim \theta \lesssim 40^\circ}$. The spirals weaken when ${\theta \gtrsim 40^\circ}$.}
    \label{fig:lowMass_misalignment}
\end{figure}

Whereas in the $0.02M_\odot$ disc, the negligible mass of the disc compared to the binary means that the total angular momentum vector of the system is effectively that of the binary. Compared to the high mass disc, the angular momentum vector of the $0.02M_\odot$ disc is further away from the total angular momentum vector of the system. Hence, it takes longer for the disc to realign. This is evident from Figure \ref{fig:lowMass_misalignment} which plots the evolution of the relative disc misalignment, $\theta$.  While the inner and outer discs are misaligned, the precession of the inner disc results in multiple cycles where the relative disc misalignment oscillates between high  (${\theta \gtrsim 40^\circ}$) and moderate $({10^\circ \lesssim \theta \lesssim 40^\circ})$ values. These cycles eventually end as the disc realigns. The leading spirals are always clearly visible (at $6, 12, 18, 21, 24$ orbits in Fig \ref{fig:lowMass}) when the inner and outer disc is moderately misaligned. When ${\theta \gtrsim 40^\circ}$ and the connection between the inner and outer disc weakens, the spirals also gradually weaken over time as seen in Fig \ref{fig:lowMass} at $3, 9, 15$ orbits.

\begin{figure*}
    \centering
    \includegraphics[width=\linewidth]{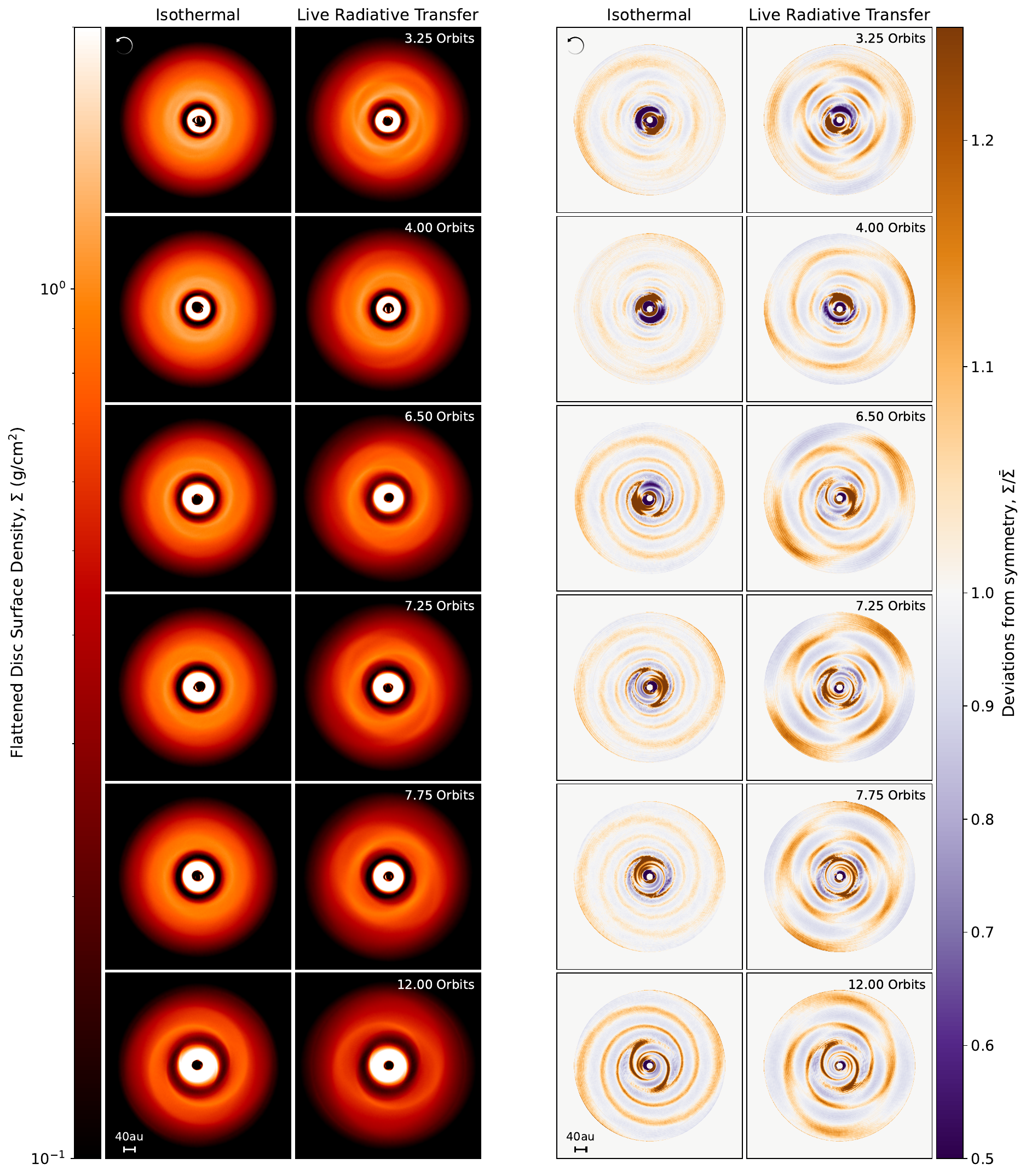}
    \caption{A comparison of the evolution of a vertically isothermal $0.02M_\odot$ circumbinary disc with an identical disc which has been modelled using live radiative transfer. The left panels show the deprojected surface density evolution. The right panel show the asymmetric structures extracted after dividing by the azimuthally averaged surface density. While leading spiral arms are formed in both simulations, there are differences in their evolution. In the isothermal simulation, the disc structure alternates between clearly visible leading spirals, and where they have weakened over time. However, with live radiative transfer, the disc structure alternates between trailing and leading spiral arms.}
    \label{fig:iso_vs_RT}
\end{figure*}

\begin{figure}
    \centering
    \includegraphics[width=\linewidth]{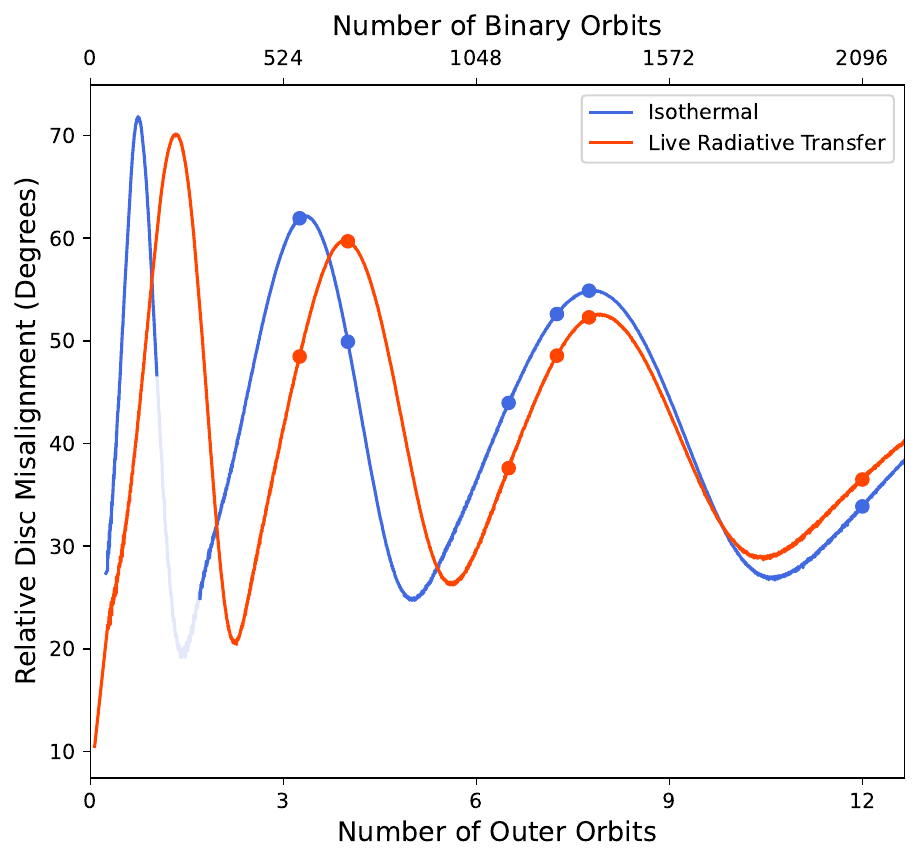}
    \caption{A comparison of the evolution of the relative disc misalignment, $\theta$, between a vertically isothermal disc and a disc modelled using live radiative transfer. The markers represent the snapshots plotted in Figure \ref{fig:iso_vs_RT}. The semi-transparent line shows when the inner disc is impacted by transient features between $1-2$ orbits. The evolution of $\theta$ is very similar for both simulations. Hence, the dynamics of the inner disc alone cannot explain the different structures seen in Fig \ref{fig:iso_vs_RT}.}
    \label{fig:iso_vs_RT_misalignment}
\end{figure}

\begin{figure}
    \centering
    \includegraphics[width=\linewidth]{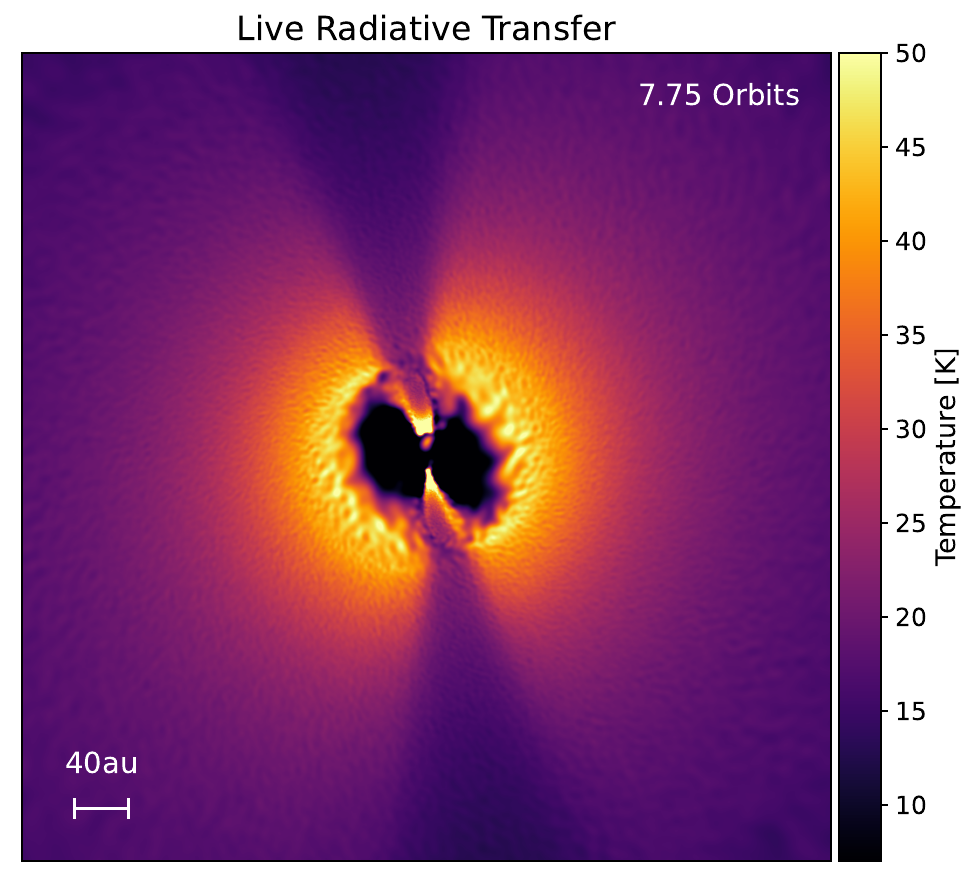}
    \caption{Cross-sectional slice of the temperature structure of the $0.02M_\odot$ disc. Using live radiative transfer results in an accurate temperature profile with azimuthal temperature differences due to the shadows cast by the misaligned inner disc.}
    \label{fig:shadow}
\end{figure}

\subsubsection{The Impact of Shadows}

Figure \ref{fig:iso_vs_RT} compares the evolution of a vertically isothermal disc with a disc using live radiative transfer to determine the disc temperatures. To extract the asymmetric structures, we first calculate the radial tilt profile of the disc,
\begin{equation}
    \beta = \cos^{-1} (\vb*{\hat{\ell}}_z),
\end{equation}
using the $z$-component of the unit angular momentum, $\vb*{\hat{\ell}}$, to unwarp the disc such that the entire disc lies flat along the $z$-axis. The transformation to flatten the disc is given by
\begin{equation}
    \begin{bmatrix} x' \\ y' \\ z' \end{bmatrix} =
    \begin{bmatrix}
    \cos(-\beta) & 0 & \sin(-\beta) \\
    0 & 1 & 0 \\
    -\sin(-\beta) & 0 & \cos(-\beta)
    \end{bmatrix}
    \begin{bmatrix} x \\ y \\ z \end{bmatrix}.
\end{equation}
The flattened discs are shown in the left panels which show the surface density evolution. Using the flattened discs, we then calculate the azimuthally averaged surface density, $\bar{\Sigma}$. Finally, to view the spiral structures more clearly, we divide the flattened surface density by $\bar{\Sigma}$. The deviations from symmetry are shown in the right panels.

It is evident that the evolution of the $0.02M\odot$ disc with live radiative transfer is different compared to the vertically isothermal disc. In addition to forming leading spiral arms, the disc also forms trailing spiral arms with live radiative transfer. Both the leading and trailing spiral arms can also interfere with each other resulting in the disc evolution becoming complicated.

Figure \ref{fig:iso_vs_RT_misalignment} compares the evolution of the relative disc misalignment, $\theta$, between both simulations. At moderate relative disc misalignments, the two simulations are similar. At 6.5 and 12 orbits, when ${\theta \lesssim 40^\circ}$, leading spiral arms are clearly visible in both simulations.  

It is at high relative disc misalignments when the two simulations differ. At 7.25 and 7.75 orbits, when ${\theta \gtrsim 40^\circ}$, the leading spirals in the vertically isothermal simulation simply weakens. At 3.25 and 4 orbits, the relative disc misalignment is even larger and the leading spirals have disappeared completely. However, in the simulation with live radiative transfer, trailing spiral arms begin to form at these times. At 3.25 and 7.25 orbits, trailing spirals start forming as the disc enters the high relative disc misalignment regime (${\theta \gtrsim 40^\circ}$). In this phase, both leading and trailing spirals coexist resulting in the complex structures visible in Fig \ref{fig:iso_vs_RT}. As $\theta$ continues to increase, the leading spirals weaken, and only trailing spirals remain. This is most easily seen at 4 and 7.75 orbits. The trailing spirals do not last long. They disappear when leading spirals begin to reform when the inner and outer disc reconnect at ${\theta \lesssim 40^\circ}$.

Fig \ref{fig:iso_vs_RT_misalignment} also shows that the evolution of the relative disc misalignment is similar between both simulations. Hence, unlike the leading spirals, the interactions between the inner and outer disc alone cannot explain the origin of the trailing spirals. 
The main difference between the two simulations is the temperature structure. In the vertically isothermal simulation, the temperature structure is axisymmetric and does not evolve with time. However, with live radiative transfer, the misaligned inner disc casts two shadows resulting in time-dependent azimuthal temperature differences. Figure \ref{fig:shadow} shows the impact of the misaligned inner disc on the temperature structure. The origin of the trailing spirals coincide with the location of the shadows. The trailing spirals seen in this work are similar to previous works which found that shadows can drive trailing spirals with zero pattern speed \citep{2024Zhang}.

\section{Discussion}
\label{sec:Discussion}

\subsection{Previous Misaligned Circumbinary discs}

The leading spiral arms visible in this work are due to interactions of the misaligned inner disc with the outer disc. Hence, it is perhaps surprising that similar features are not visible in the numerous previous studies of misaligned circumbinary discs. There are a few reasons why leading spirals were less visible in previous work. The precession of the inner disc impacts the evolution of the leading spiral arms. The resolution of the disc and transient features also have an impact.

The impact of disc precession on the strength of the leading spirals can be seen in the lower mass disc where the slower realignment allows the inner disc to precess continuously for a longer duration. As the inner disc precesses, the relative disc misalignment between the inner and outer disc oscillates between low and high values, causing the leading spirals to alternate between being clearly visible, and where they have weakened over time. The precession rate is dependent on binary parameters such as the semi-major axis and mass ratio \citep{2010Farago,2018Lubow,2022Martin}. The binary parameters can have a vast range of values leading to very different precession rates which could disrupt the formation of leading spirals. 

The resolution of the simulations can also impact the visibility of leading spiral arms. The amount of mass accreted is resolution dependent \citep{2015Nealon}, which is particularly relevant for warped discs as it can impact the alignment and precession of the disc. Additionally, at lower resolution, the inner disc is impacted by transient features (see Figure \ref{fig:resolution}). The formation of leading spirals is delayed until the transient features have disappeared and the disc has settled. Thus, if the resolution is too low and the simulation has not been run for long enough (${>}1000$ binary orbits), the formation of leading spiral arms would be missed.

A combination of these factors could explain why leading spiral arms are not seen as often. However, close inspection of previous studies of misaligned circumbinary simulations shows that these spirals have been captured previously. For example, the $i=60^\circ$ simulation in Figure 4 in \citet{2018Aly}, the $i=45^\circ$ simulation in Figure 6 in \citet{2020Hirsh}, and panel 2 of Figure 4 in \citet{2024Rabago}. While the disc setups are not identical, the leading spiral arms are visible with similar initial binary misalignments as in this work.

\subsection{Previous Shadow-Driven Spirals}

The presence of shadows have been shown to launch trailing spiral arms \citep{2016Montesinos,2019bCuello,2024Zhang}. While the live radiative transfer simulations in this work are able to self-consistently model shadows, the spirals formed differ from previous work. In this work, the spirals are usually leading instead of trailing.

The cause of this discrepancy is in how the dynamics of the misaligned inner disc are modelled. The shadows in \cite{2016Montesinos,2019bCuello} are prescribed. \cite{2024Zhang} do not model the hydrodynamics of the inner disc. Whereas in our simulations, the coupling of \textsc{Phantom} and \textsc{mcfost} means that we model the hydrodynamics of the inner disc with \textsc{Phantom}, and shadowing effects are accounted for with \textsc{mcfost}. The inclusion of the misaligned inner disc results in an additional mechanism for launching spirals (see Figure \ref{fig:isoAltView}). These spirals are leading, and separate from shadow-driven spirals seen in previous studies. It is only at high relative misalignments, when the inner disc is no longer connected to the outer disc, that trailing spiral arms due to shadows are visible. The dynamical interaction between the inner and outer disc overwhelms the shadows. Thus, our results are consistent with \cite{2016Montesinos,2019bCuello,2024Zhang,2024Su}.

While \cite{2020Nealon} also modelled a misaligned circumbinary disc with \textsc{Phantom} coupled with \textsc{mcfost}, they did not see either trailing or leading spiral arms. This is likely due to the radial extent of the disc (13.3au) used in their work being much narrower than the radial extent of the spirals (${\sim}50$au) in this work. Whereas the disc size used here is similar to \cite{2016Montesinos,2019bCuello,2024Zhang}.

\subsection{Observational Implications}

The results in this work show that a misalignment between the inner and outer disc provides a mechanism to simultaneously form large scale spiral arms and a gap. Notably, the spirals can be either trailing or leading depending on whether the disc is broken or not. Unlike spirals driven by gravitational instability, this mechanism does not require high mass discs. It also does not require the presence of planets to open up a gap. 

Other studies have also found spiral structures in misaligned discs in binary systems such as HD 142527 \citep{2018bPrice,2024Nowak}. However, the mechanism described in this work is different. Compared to systems like HD 142527, the difference is due to the location of the inner disc relative to the stars. In HD 142527, the inner disc is a circumprimary disc, whereas here it is a circumbinary disc. Additionally, there is a complete break between the inner and outer disc in HD 142527. As shown in \citet{2018bPrice}, these differences result in a distinct origin and evolution of the spirals. When the inner disc gets disrupted, the spirals are unaffected as they are driven by the interaction of the secondary star with the outer disc. In contrast, in our work, the spirals are formed due to the interaction of the inner disc with the outer disc. Hence, disruption of the inner disc also causes the spiral structures to disappear (see Figure \ref{fig:resolution}), highlighting the different origin.

Larger grains trace the dust continuum, whereas scattered light observations trace smaller grains. In the lower-mass disc, larger millimeter-sized dust grains correspond to Stokes numbers of ${St\sim 0.1}$, implying the dust is weakly coupled to the gas. Thus, dedicated dust simulations are required for accurate observational predictions of the continuum which are beyond the scope of this study. It is therefore unclear if the millimeter dust would still form spiral arms, or result in different structures as seen in \citet{2024Aly}.  

However in the $0.2M_\odot$ disc, millimeter-sized dust grains have lower Stokes numbers of  $St\sim 10^{-2}$, implying that the dust is more strongly coupled to the gas. Thus, we can make reasonable predictions assuming perfect coupling.  Hence, it is possible the leading spiral arms could be seen in dust continuum observations. Due to the misalignment between the inner and outer disc, these spiral would be accompanied by shadows in scattered light observations \citep{2015Marino,2018Facchini,2019Nealon}.

\section{Conclusions}
\label{sec:conclusion}

We use 3D hydrodynamics to investigate the evolution of misaligned circumbinary discs. Our main findings are:

\begin{enumerate}
    \item In a nearly broken disc, the misaligned inner disc remains connected to the outer disc at two nodes from which spirals are launched. Unlike typical mechanisms for launching spirals, these spirals do not rotate with the disc, and are leading instead of trailing.
    \item If the disc completely breaks into two separate discs, or if the disc is aligned, the leading spirals disappear since the inner and outer disc are no longer connected at two narrow nodes.
    \item We also perform simulations with live radiative transfer and self-gravity which demonstrate that the formation of leading spirals is independent of the disc physics. The relative misalignment between the inner and outer disc drives the evolution of the leading spiral arms.
    \item For the first time, we also recover shadow driven trailing spiral arms in a dynamical calculation with a broken disc; consistent with predictions from previous works. However, they are only visible at high relative misalignments when the inner disc is disconnected from the outer disc. At lower misalignments, we find that these spirals interact with the leading spirals that are driven by the interaction with the inner disc.
\end{enumerate}

\section*{Acknowledgements}
We thank the reviewer, Dhruv Muley for their comments that benefitted this work. SR would like to thank Hossam Aly, Cathie Clarke, Amelia Cordwell, Callum Fairbairn, Giuseppe Lodato, Rebecca Martin, and Alison Young for interesting discussions. SR and RA acknowledge support funding from the Science \& Technology Facilities Council (STFC) through Consolidated Grant ST/W000857/1. FM acknowledges a Royal Society Dorothy Hodgkin Fellowship. RN acknowledges UKRI/EPSRC support via a Stephen Hawking Fellowship (EP/T017287/1). We used the ALICE High Performance Computing Facility and the DiRAC Data Intensive service (DIaL), managed by the University of Leicester Research Computing Service on behalf of the STFC DiRAC HPC Facility (www.dirac.ac.uk). The DiRAC service at Leicester was funded by BEIS, UKRI and STFC capital funding and STFC operations grants. DiRAC is part of the UKRI Digital Research Infrastructure.

\section*{Data Availability}

Videos and select snapshots of the simulations along with the data to recreate any of the simulations or the figures in the paper are available at 
\url{https://doi.org/10.5281/zenodo.16750300}. We utilised the following public software:\\

\begin{tabular}{ll}
\hspace{-0.52cm}\textsc{Phantom}
&\url{https://github.com/danieljprice/phantom}\\
& \citep{2018Price} \\
\hspace{-0.52cm}\textsc{mcfost} &\url{https://github.com/cpinte/mcfost}\\
& \citep{2006Pinte,2009Pinte} \\
\hspace{-0.52cm}\textsc{Splash}
&\url{https://github.com/danieljprice/splash}\\ & \citep{2007Price} \\
\hspace{-0.52cm}\textsc{Sarracen} &\url{https://github.com/ttricco/sarracen}\\
& \citep{sarracen}
\end{tabular}



\bibliographystyle{mnras.bst}
\bibliography{LeadingSpirals.bib} 

\begin{thebibliography}{}
\makeatletter
\relax
\def\mn@urlcharsother{\let\do\@makeother \do\$\do\&\do\#\do\^\do\_\do\%\do\~}
\def\mn@doi{\begingroup\mn@urlcharsother \@ifnextchar [ {\mn@doi@} {\mn@doi@[]}}
\def\mn@doi@[#1]#2{\def\@tempa{#1}\ifx\@tempa\@empty \href {http://dx.doi.org/#2} {doi:#2}\else \href {http://dx.doi.org/#2} {\color{violet}#1}\fi \endgroup}
\def\mn@eprint#1#2{\mn@eprint@#1:#2::\@nil}
\def\mn@eprint@arXiv#1{\href {http://arxiv.org/abs/#1} {{\tt arXiv:#1}}}
\def\mn@eprint@dblp#1{\href {http://dblp.uni-trier.de/rec/bibtex/#1.xml} {dblp:#1}}
\def\mn@eprint@#1:#2:#3:#4\@nil{\def\@tempa {#1}\def\@tempb {#2}\def\@tempc {#3}\ifx \@tempc \@empty \let \@tempc \@tempb \let \@tempb \@tempa \fi \ifx \@tempb \@empty \def\@tempb {arXiv}\fi \@ifundefined {mn@eprint@\@tempb}{\@tempb:\@tempc}{\expandafter \expandafter \csname mn@eprint@\@tempb\endcsname \expandafter{\@tempc}}}

\bibitem[\protect\citeauthoryear{{ALMA Partnership} et~al.,}{{ALMA Partnership} et~al.}{2015}]{2015ALMA}
{ALMA Partnership} et~al., 2015, \mn@doi [\apj] {10.1088/2041-8205/808/1/L3}, \href {https://ui.adsabs.harvard.edu/abs/2015ApJ...808L...3A} {808, L3}

\bibitem[\protect\citeauthoryear{{Aly}, {Lodato}  \& {Cazzoletti}}{{Aly} et~al.}{2018}]{2018Aly}
{Aly} H.,  {Lodato} G.,   {Cazzoletti} P.,  2018, \mn@doi [\mnras] {10.1093/mnras/sty2179}, \href {https://ui.adsabs.harvard.edu/abs/2018MNRAS.480.4738A} {480, 4738}

\bibitem[\protect\citeauthoryear{{Aly}, {Nealon}  \& {Gonzalez}}{{Aly} et~al.}{2024}]{2024Aly}
{Aly} H.,  {Nealon} R.,   {Gonzalez} J.-F.,  2024, \mn@doi [\mnras] {10.1093/mnras/stad3494}, \href {https://ui.adsabs.harvard.edu/abs/2024MNRAS.527.4777A} {527, 4777}

\bibitem[\protect\citeauthoryear{{Andrews} et~al.,}{{Andrews} et~al.}{2018}]{2018Andrews}
{Andrews} S.~M.,  et~al., 2018, \mn@doi [\apj] {10.3847/2041-8213/aaf741}, \href {https://ui.adsabs.harvard.edu/abs/2018ApJ...869L..41A} {869, L41}

\bibitem[\protect\citeauthoryear{{Ballabio}, {Nealon}, {Alexander}, {Cuello}, {Pinte}  \& {Price}}{{Ballabio} et~al.}{2021}]{2021Ballabio}
{Ballabio} G.,  {Nealon} R.,  {Alexander} R.~D.,  {Cuello} N.,  {Pinte} C.,   {Price} D.~J.,  2021, \mn@doi [\mnras] {10.1093/mnras/stab922}, \href {https://ui.adsabs.harvard.edu/abs/2021MNRAS.504..888B} {504, 888}

\bibitem[\protect\citeauthoryear{Bate}{Bate}{2018}]{2018Bate}
Bate M.~R.,  2018, \mn@doi [\mnras] {10.1093/mnras/sty169}, 475, 5618

\bibitem[\protect\citeauthoryear{{Bate}, {Bonnell}  \& {Price}}{{Bate} et~al.}{1995}]{1995Bate}
{Bate} M.~R.,  {Bonnell} I.~A.,   {Price} N.~M.,  1995, \mn@doi [\mnras] {10.1093/mnras/277.2.362}, \href {https://ui.adsabs.harvard.edu/abs/1995MNRAS.277..362B} {277, 362}

\bibitem[\protect\citeauthoryear{{Benisty} et~al.,}{{Benisty} et~al.}{2015}]{2015Benisty}
{Benisty} M.,  et~al., 2015, \mn@doi [\aap] {10.1051/0004-6361/201526011}, \href {https://ui.adsabs.harvard.edu/abs/2015A&A...578L...6B} {578, L6}

\bibitem[\protect\citeauthoryear{{Benisty} et~al.,}{{Benisty} et~al.}{2017}]{2017Benisty}
{Benisty} M.,  et~al., 2017, \mn@doi [\aap] {10.1051/0004-6361/201629798}, \href {https://ui.adsabs.harvard.edu/abs/2017A&A...597A..42B} {597, A42}

\bibitem[\protect\citeauthoryear{{Bohn} et~al.,}{{Bohn} et~al.}{2022}]{2022Bohn}
{Bohn} A.~J.,  et~al., 2022, \mn@doi [\aap] {10.1051/0004-6361/202142070}, \href {https://ui.adsabs.harvard.edu/abs/2022A&A...658A.183B} {658, A183}

\bibitem[\protect\citeauthoryear{{Borchert}, {Price}, {Pinte}  \& {Cuello}}{{Borchert} et~al.}{2022a}]{2022aBorchert}
{Borchert} E. M.~A.,  {Price} D.~J.,  {Pinte} C.,   {Cuello} N.,  2022a, \mn@doi [\mnras] {10.1093/mnrasl/slab123}, \href {https://ui.adsabs.harvard.edu/abs/2022MNRAS.510L..37B} {510, L37}

\bibitem[\protect\citeauthoryear{{Borchert} et~al.}{{Borchert} et~al.}{2022b}]{2022bBorchert}
{Borchert} E. M.~A.,  et~al., 2022b, \mn@doi [\mnras] {10.1093/mnras/stac2872}, \href {https://ui.adsabs.harvard.edu/abs/2022MNRAS.517.4436B} {517, 4436}

\bibitem[\protect\citeauthoryear{{Chiang} \& {Goldreich}}{{Chiang} \& {Goldreich}}{1997}]{1997Chiang}
{Chiang} E.~I.,  {Goldreich} P.,  1997, \mn@doi [\apj] {10.1086/304869}, \href {https://ui.adsabs.harvard.edu/abs/1997ApJ...490..368C} {490, 368}

\bibitem[\protect\citeauthoryear{{Cuello}, {Montesinos}, {Stammler}, {Louvet}  \& {Cuadra}}{{Cuello} et~al.}{2019}]{2019bCuello}
{Cuello} N.,  {Montesinos} M.,  {Stammler} S.~M.,  {Louvet} F.,   {Cuadra} J.,  2019, \mn@doi [\aap] {10.1051/0004-6361/201731732}, \href {https://ui.adsabs.harvard.edu/abs/2019A&A...622A..43C} {622, A43}

\bibitem[\protect\citeauthoryear{{Cullen} \& {Dehnen}}{{Cullen} \& {Dehnen}}{2010}]{2010Cullen}
{Cullen} L.,  {Dehnen} W.,  2010, \mn@doi [\mnras] {10.1111/j.1365-2966.2010.17158.x}, \href {https://ui.adsabs.harvard.edu/abs/2010MNRAS.408..669C} {408, 669}

\bibitem[\protect\citeauthoryear{{Facchini}, {Lodato}  \& {Price}}{{Facchini} et~al.}{2013}]{2013Facchini}
{Facchini} S.,  {Lodato} G.,   {Price} D.~J.,  2013, \mn@doi [\mnras] {10.1093/mnras/stt877}, \href {https://ui.adsabs.harvard.edu/abs/2013MNRAS.433.2142F} {433, 2142}

\bibitem[\protect\citeauthoryear{{Facchini}, {Juh{\'a}sz}  \& {Lodato}}{{Facchini} et~al.}{2018}]{2018Facchini}
{Facchini} S.,  {Juh{\'a}sz} A.,   {Lodato} G.,  2018, \mn@doi [\mnras] {10.1093/mnras/stx2523}, \href {https://ui.adsabs.harvard.edu/abs/2018MNRAS.473.4459F} {473, 4459}

\bibitem[\protect\citeauthoryear{{Farago} \& {Laskar}}{{Farago} \& {Laskar}}{2010}]{2010Farago}
{Farago} F.,  {Laskar} J.,  2010, \mn@doi [\mnras] {10.1111/j.1365-2966.2009.15711.x}, \href {https://ui.adsabs.harvard.edu/abs/2010MNRAS.401.1189F} {401, 1189}

\bibitem[\protect\citeauthoryear{{Forgan}, {Ilee}  \& {Meru}}{{Forgan} et~al.}{2018}]{2018bForgan}
{Forgan} D.~H.,  {Ilee} J.~D.,   {Meru} F.,  2018, \mn@doi [\apjl] {10.3847/2041-8213/aac7c9}, \href {https://ui.adsabs.harvard.edu/abs/2018ApJ...860L...5F} {860, L5}

\bibitem[\protect\citeauthoryear{{Garufi} et~al.,}{{Garufi} et~al.}{2024}]{2024Garufi}
{Garufi} A.,  et~al., 2024, \mn@doi [\aap] {10.1051/0004-6361/202347586}, \href {https://ui.adsabs.harvard.edu/abs/2024A&A...685A..53G} {685, A53}

\bibitem[\protect\citeauthoryear{Harris \& Tricco}{Harris \& Tricco}{2023}]{sarracen}
Harris A.,  Tricco T.~S.,  2023, \mn@doi [Journal of Open Source Software] {10.21105/joss.05263}, 8, 5263

\bibitem[\protect\citeauthoryear{{Hirsh}, {Price}, {Gonzalez}, {Ubeira-Gabellini}  \& {Ragusa}}{{Hirsh} et~al.}{2020}]{2020Hirsh}
{Hirsh} K.,  {Price} D.~J.,  {Gonzalez} J.-F.,  {Ubeira-Gabellini} M.~G.,   {Ragusa} E.,  2020, \mn@doi [\mnras] {10.1093/mnras/staa2536}, \href {https://ui.adsabs.harvard.edu/abs/2020MNRAS.498.2936H} {498, 2936}

\bibitem[\protect\citeauthoryear{{Huang} et~al.,}{{Huang} et~al.}{2018a}]{2018Huang}
{Huang} J.,  et~al., 2018a, \mn@doi [\apjl] {10.3847/2041-8213/aaf740}, \href {https://ui.adsabs.harvard.edu/abs/2018ApJ...869L..42H} {869, L42}

\bibitem[\protect\citeauthoryear{{Huang} et~al.,}{{Huang} et~al.}{2018b}]{2018bHuang}
{Huang} J.,  et~al., 2018b, \mn@doi [\apjl] {10.3847/2041-8213/aaf7a0}, \href {https://ui.adsabs.harvard.edu/abs/2018ApJ...869L..43H} {869, L43}

\bibitem[\protect\citeauthoryear{{Keppler} et~al.,}{{Keppler} et~al.}{2020}]{2020Keppler}
{Keppler} M.,  et~al., 2020, \mn@doi [\aap] {10.1051/0004-6361/202038032}, \href {https://ui.adsabs.harvard.edu/abs/2020A&A...639A..62K} {639, A62}

\bibitem[\protect\citeauthoryear{{King}, {Lubow}, {Ogilvie}  \& {Pringle}}{{King} et~al.}{2005}]{2005King}
{King} A.~R.,  {Lubow} S.~H.,  {Ogilvie} G.~I.,   {Pringle} J.~E.,  2005, \mn@doi [\mnras] {10.1111/j.1365-2966.2005.09378.x}, \href {https://ui.adsabs.harvard.edu/abs/2005MNRAS.363...49K} {363, 49}

\bibitem[\protect\citeauthoryear{{Kley} \& {Nelson}}{{Kley} \& {Nelson}}{2012}]{2012Kley}
{Kley} W.,  {Nelson} R.~P.,  2012, \mn@doi [\araa] {10.1146/annurev-astro-081811-125523}, \href {https://ui.adsabs.harvard.edu/abs/2012ARA&A..50..211K} {50, 211}

\bibitem[\protect\citeauthoryear{{Lodato} \& {Pringle}}{{Lodato} \& {Pringle}}{2007}]{2007Lodato}
{Lodato} G.,  {Pringle} J.~E.,  2007, \mn@doi [\mnras] {10.1111/j.1365-2966.2007.12332.x}, \href {https://ui.adsabs.harvard.edu/abs/2007MNRAS.381.1287L} {381, 1287}

\bibitem[\protect\citeauthoryear{{Long} et~al.,}{{Long} et~al.}{2018}]{2018Long}
{Long} F.,  et~al., 2018, \mn@doi [\apj] {10.3847/1538-4357/aae8e1}, \href {https://ui.adsabs.harvard.edu/abs/2018ApJ...869...17L} {869, 17}

\bibitem[\protect\citeauthoryear{{Lubow} \& {Martin}}{{Lubow} \& {Martin}}{2018}]{2018Lubow}
{Lubow} S.~H.,  {Martin} R.~G.,  2018, \mn@doi [\mnras] {10.1093/mnras/stx2643}, \href {https://ui.adsabs.harvard.edu/abs/2018MNRAS.473.3733L} {473, 3733}

\bibitem[\protect\citeauthoryear{{Marino}, {Perez}  \& {Casassus}}{{Marino} et~al.}{2015}]{2015Marino}
{Marino} S.,  {Perez} S.,   {Casassus} S.,  2015, \mn@doi [\apjl] {10.1088/2041-8205/798/2/L44}, \href {https://ui.adsabs.harvard.edu/abs/2015ApJ...798L..44M} {798, L44}

\bibitem[\protect\citeauthoryear{{Martin}, {Lepp}, {Lubow}, {Kenworthy}, {Kennedy}  \& {Vallet}}{{Martin} et~al.}{2022}]{2022Martin}
{Martin} R.~G.,  {Lepp} S.,  {Lubow} S.~H.,  {Kenworthy} M.~A.,  {Kennedy} G.~M.,   {Vallet} D.,  2022, \mn@doi [\apjl] {10.3847/2041-8213/ac54b4}, \href {https://ui.adsabs.harvard.edu/abs/2022ApJ...927L..26M} {927, L26}

\bibitem[\protect\citeauthoryear{{Meru} et~al.}{{Meru} et~al.}{2017}]{2017Meru}
{Meru} F.,  et~al., 2017, \mn@doi [\apjl] {10.3847/2041-8213/aa6837}, \href {https://ui.adsabs.harvard.edu/abs/2017ApJ...839L..24M} {839, L24}

\bibitem[\protect\citeauthoryear{{Montesinos}, {Perez}, {Casassus}, {Marino}, {Cuadra}  \& {Christiaens}}{{Montesinos} et~al.}{2016}]{2016Montesinos}
{Montesinos} M.,  {Perez} S.,  {Casassus} S.,  {Marino} S.,  {Cuadra} J.,   {Christiaens} V.,  2016, \mn@doi [\apjl] {10.3847/2041-8205/823/1/L8}, \href {https://ui.adsabs.harvard.edu/abs/2016ApJ...823L...8M} {823, L8}

\bibitem[\protect\citeauthoryear{{Muro-Arena} et~al.,}{{Muro-Arena} et~al.}{2020}]{2020Muro-Arena}
{Muro-Arena} G.~A.,  et~al., 2020, \mn@doi [\aap] {10.1051/0004-6361/201936509}, \href {https://ui.adsabs.harvard.edu/abs/2020A&A...635A.121M} {635, A121}

\bibitem[\protect\citeauthoryear{{Nealon}, {Price}  \& {Nixon}}{{Nealon} et~al.}{2015}]{2015Nealon}
{Nealon} R.,  {Price} D.~J.,   {Nixon} C.~J.,  2015, \mn@doi [\mnras] {10.1093/mnras/stv014}, \href {https://ui.adsabs.harvard.edu/abs/2015MNRAS.448.1526N} {448, 1526}

\bibitem[\protect\citeauthoryear{{Nealon}, {Pinte}, {Alexander}, {Mentiplay}  \& {Dipierro}}{{Nealon} et~al.}{2019}]{2019Nealon}
{Nealon} R.,  {Pinte} C.,  {Alexander} R.,  {Mentiplay} D.,   {Dipierro} G.,  2019, \mn@doi [\mnras] {10.1093/mnras/stz346}, \href {https://ui.adsabs.harvard.edu/abs/2019MNRAS.484.4951N} {484, 4951}

\bibitem[\protect\citeauthoryear{{Nealon}, {Cuello}  \& {Alexander}}{{Nealon} et~al.}{2020a}]{2020aNealon}
{Nealon} R.,  {Cuello} N.,   {Alexander} R.,  2020a, \mn@doi [\mnras] {10.1093/mnras/stz3186}, \href {https://ui.adsabs.harvard.edu/abs/2020MNRAS.491.4108N} {491, 4108}

\bibitem[\protect\citeauthoryear{{Nealon}, {Price}  \& {Pinte}}{{Nealon} et~al.}{2020b}]{2020Nealon}
{Nealon} R.,  {Price} D.~J.,   {Pinte} C.,  2020b, \mn@doi [\mnras] {10.1093/mnrasl/slaa026}, \href {https://ui.adsabs.harvard.edu/abs/2020MNRAS.493L.143N} {493, L143}

\bibitem[\protect\citeauthoryear{{Nowak}, {Rowther}, {Lacour}, {Meru}, {Nealon}  \& {Price}}{{Nowak} et~al.}{2024}]{2024Nowak}
{Nowak} M.,  {Rowther} S.,  {Lacour} S.,  {Meru} F.,  {Nealon} R.,   {Price} D.~J.,  2024, \mn@doi [\aap] {10.1051/0004-6361/202347748}, \href {https://ui.adsabs.harvard.edu/abs/2024A&A...683A...6N} {683, A6}

\bibitem[\protect\citeauthoryear{{Paneque-Carre{\~n}o} et~al.,}{{Paneque-Carre{\~n}o} et~al.}{2021}]{2021Paneque}
{Paneque-Carre{\~n}o} T.,  et~al., 2021, \mn@doi [\apj] {10.3847/1538-4357/abf243}, \href {https://ui.adsabs.harvard.edu/abs/2021ApJ...914...88P} {914, 88}

\bibitem[\protect\citeauthoryear{{P{\'e}rez} et~al.,}{{P{\'e}rez} et~al.}{2016}]{2016Perez}
{P{\'e}rez} L.~M.,  et~al., 2016, \mn@doi [Science] {10.1126/science.aaf8296}, \href {https://ui.adsabs.harvard.edu/abs/2016Sci...353.1519P} {353, 1519}

\bibitem[\protect\citeauthoryear{{Pinte}, {M{\'e}nard}, {Duch{\^e}ne}  \& {Bastien}}{{Pinte} et~al.}{2006}]{2006Pinte}
{Pinte} C.,  {M{\'e}nard} F.,  {Duch{\^e}ne} G.,   {Bastien} P.,  2006, \mn@doi [\aap] {10.1051/0004-6361:20053275}, \href {https://ui.adsabs.harvard.edu/abs/2006A&A...459..797P} {459, 797}

\bibitem[\protect\citeauthoryear{{Pinte} et~al.}{{Pinte} et~al.}{2009}]{2009Pinte}
{Pinte} C.,  et~al., 2009, \mn@doi [\aap] {10.1051/0004-6361/200811555}, \href {https://ui.adsabs.harvard.edu/abs/2009A&A...498..967P} {498, 967}

\bibitem[\protect\citeauthoryear{{Price}}{{Price}}{2007}]{2007Price}
{Price} D.~J.,  2007, \mn@doi [\pasa] {10.1071/AS07022}, \href {https://ui.adsabs.harvard.edu/abs/2007PASA...24..159P} {24, 159}

\bibitem[\protect\citeauthoryear{{Price} et~al.,}{{Price} et~al.}{2018a}]{2018Price}
{Price} D.~J.,  et~al., 2018a, \mn@doi [\pasa] {10.1017/pasa.2018.25}, \href {https://ui.adsabs.harvard.edu/abs/2018PASA...35...31P} {35, e031}

\bibitem[\protect\citeauthoryear{{Price} et~al.,}{{Price} et~al.}{2018b}]{2018bPrice}
{Price} D.~J.,  et~al., 2018b, \mn@doi [\mnras] {10.1093/mnras/sty647}, \href {https://ui.adsabs.harvard.edu/abs/2018MNRAS.477.1270P} {477, 1270}

\bibitem[\protect\citeauthoryear{{Rabago}, {Zhu}, {Lubow}  \& {Martin}}{{Rabago} et~al.}{2024}]{2024Rabago}
{Rabago} I.,  {Zhu} Z.,  {Lubow} S.,   {Martin} R.~G.,  2024, \mn@doi [\mnras] {10.1093/mnras/stae1787}, \href {https://ui.adsabs.harvard.edu/abs/2024MNRAS.533..360R} {533, 360}

\bibitem[\protect\citeauthoryear{{Rowther}, {Nealon}  \& {Meru}}{{Rowther} et~al.}{2022}]{2022Rowther}
{Rowther} S.,  {Nealon} R.,   {Meru} F.,  2022, \mn@doi [\apj] {10.3847/1538-4357/ac3975}, \href {https://ui.adsabs.harvard.edu/abs/2022ApJ...925..163R} {925, 163}

\bibitem[\protect\citeauthoryear{{Rowther}, {Price}, {Pinte}, {Nealon}, {Meru}  \& {Alexander}}{{Rowther} et~al.}{2024}]{2024bRowther}
{Rowther} S.,  {Price} D.~J.,  {Pinte} C.,  {Nealon} R.,  {Meru} F.,   {Alexander} R.,  2024, \mn@doi [\mnras] {10.1093/mnras/stae2167}, \href {https://ui.adsabs.harvard.edu/abs/2024MNRAS.534.2277R} {534, 2277}

\bibitem[\protect\citeauthoryear{{Sakai}, {Hanawa}, {Zhang}, {Higuchi}, {Ohashi}, {Oya}  \& {Yamamoto}}{{Sakai} et~al.}{2019}]{2019Sakai}
{Sakai} N.,  {Hanawa} T.,  {Zhang} Y.,  {Higuchi} A.~E.,  {Ohashi} S.,  {Oya} Y.,   {Yamamoto} S.,  2019, \mn@doi [\nat] {10.1038/s41586-018-0819-2}, \href {https://ui.adsabs.harvard.edu/abs/2019Natur.565..206S} {565, 206}

\bibitem[\protect\citeauthoryear{Segura-Cox et~al.,}{Segura-Cox et~al.}{2020}]{Segura-Cox2020}
Segura-Cox D.~M.,  et~al., 2020, \mn@doi [Nature] {10.1038/s41586-020-2779-6}, 586, 228

\bibitem[\protect\citeauthoryear{{Shakura} \& {Sunyaev}}{{Shakura} \& {Sunyaev}}{1973}]{1973SS}
{Shakura} N.~I.,  {Sunyaev} R.~A.,  1973, \aap, \href {https://ui.adsabs.harvard.edu/abs/1973A&A....24..337S} {24, 337}

\bibitem[\protect\citeauthoryear{{Sheehan} \& {Eisner}}{{Sheehan} \& {Eisner}}{2018}]{2018Sheehan}
{Sheehan} P.~D.,  {Eisner} J.~A.,  2018, \mn@doi [\apj] {10.3847/1538-4357/aaae65}, \href {https://ui.adsabs.harvard.edu/abs/2018ApJ...857...18S} {857, 18}

\bibitem[\protect\citeauthoryear{{Siess}, {Dufour}  \& {Forestini}}{{Siess} et~al.}{2000}]{2000Siess}
{Siess} L.,  {Dufour} E.,   {Forestini} M.,  2000, \aap, \href {https://ui.adsabs.harvard.edu/abs/2000A&A...358..593S} {358, 593}

\bibitem[\protect\citeauthoryear{{Smallwood}, {Yang}, {Zhu}, {Martin}, {Dong}, {Cuello}  \& {Isella}}{{Smallwood} et~al.}{2023}]{2023Smallwood}
{Smallwood} J.~L.,  {Yang} C.-C.,  {Zhu} Z.,  {Martin} R.~G.,  {Dong} R.,  {Cuello} N.,   {Isella} A.,  2023, \mn@doi [\mnras] {10.1093/mnras/stad742}, \href {https://ui.adsabs.harvard.edu/abs/2023MNRAS.521.3500S} {521, 3500}

\bibitem[\protect\citeauthoryear{{Speedie} et~al.}{{Speedie} et~al.}{2024}]{2024Speedie}
{Speedie} J.,  et~al., 2024, \mn@doi [\nat] {10.1038/s41586-024-07877-0}, \href {https://ui.adsabs.harvard.edu/abs/2024Natur.633...58S} {633, 58}

\bibitem[\protect\citeauthoryear{{Speedie} et~al.}{{Speedie} et~al.}{2025}]{2025Speedie}
{Speedie} J.,  et~al., 2025, \mn@doi [\apjl] {10.3847/2041-8213/adb7d5}, \href {https://ui.adsabs.harvard.edu/abs/2025ApJ...981L..30S} {981, L30}

\bibitem[\protect\citeauthoryear{{Su} \& {Bai}}{{Su} \& {Bai}}{2024}]{2024Su}
{Su} Z.,  {Bai} X.-N.,  2024, \mn@doi [\apj] {10.3847/1538-4357/ad7581}, \href {https://ui.adsabs.harvard.edu/abs/2024ApJ...975..126S} {975, 126}

\bibitem[\protect\citeauthoryear{{Villenave} et~al.,}{{Villenave} et~al.}{2024}]{2024Villenave}
{Villenave} M.,  et~al., 2024, \mn@doi [\apj] {10.3847/1538-4357/ad0c4b}, \href {https://ui.adsabs.harvard.edu/abs/2024ApJ...961...95V} {961, 95}

\bibitem[\protect\citeauthoryear{{Weingartner} \& {Draine}}{{Weingartner} \& {Draine}}{2001}]{Weingartner2001}
{Weingartner} J.~C.,  {Draine} B.~T.,  2001, \mn@doi [\apj] {10.1086/318651}, \href {https://ui.adsabs.harvard.edu/abs/2001ApJ...548..296W} {548, 296}

\bibitem[\protect\citeauthoryear{{Zhang} \& {Zhu}}{{Zhang} \& {Zhu}}{2024}]{2024Zhang}
{Zhang} S.,  {Zhu} Z.,  2024, \mn@doi [\apjl] {10.3847/2041-8213/ad815f}, \href {https://ui.adsabs.harvard.edu/abs/2024ApJ...974L..38Z} {974, L38}

\makeatother
\end{thebibliography}


\appendix




\bsp	
\label{lastpage}
\end{document}